%% file: main.tex
\UseRawInputEncoding
\documentclass[preprint,review,12pt]{elsarticle}

\usepackage{amsmath}
\usepackage{amssymb}
\usepackage{amsthm}
\usepackage{graphicx}
\usepackage{float}
\usepackage{wrapfig}
\usepackage{caption}
\usepackage{subcaption}
\usepackage{placeins}
\usepackage{xcolor}

\usepackage{lineno}
\usepackage{hyperref}

\journal{}
\begin{document}
\begin{frontmatter}

\title{Stochastic Soiling Loss Models for Heliostats in Concentrating Solar Power Plants}
\author[inst1]{Giovanni Picotti}
\author[inst1]{Michael E. Cholette}
\author[inst1,inst2]{Cody B. Anderson}
\author[inst1]{Theodore A. Steinberg}
\author[inst2]{Giampaolo Manzolini}

\affiliation[inst1]{organization={School of Mechanical, Medical, and Process Engineering, Queensland University of Technology},
            addressline={2 George St.}, 
            city={Brisbane},
            postcode={4000}, 
            state={Queensland},
            country={Australia}}

\affiliation[inst2]{organization={Dipartimento di Energia, Politecnico di Milano},
            addressline={Via Lambruschini~4}, 
            postcode={20156},
            city={Milano},
            country={Italy}}

\begin{abstract}
   Reflectance losses on solar mirrors due to soiling are a significant challenge for Concentrating Solar Power (CSP) plants. Soiling losses can vary significantly from site to site --- with (absolute) reflectance losses varying from fractions of a percentage point up to several percentage points per day (pp/day), a fact that has motivated several studies in soiling predictive modelling. Yet, existing studies have so far neglected the characterization of statistical uncertainty in their parameters and predictions.  In this paper, two reflectance loss models are proposed that model uncertainty: an extension of a previously developed physical model and a simplified model. A novel uncertainty characterization enables Maximum Likelihood Estimation techniques for parameter estimation for both models, and permits the estimation of parameter (and prediction) confidence intervals. 
   
   The models are applied to data from ten soiling campaigns conducted at three Australian sites (Brisbane, Mount Isa, Wodonga). The simplified model produces high-quality predictions of soiling losses on novel data, while the semi-physical model performance is mixed. The statistical distributions of daily losses were estimated for different dust loadings. Under median conditions, the daily soiling losses for Brisbane, Mount Isa, and Wodonga are estimated as $0.53 \pm 0.66$, $0.08 \pm 0.08$, and $0.58 \pm 0.15$ pp/day, respectively. Yet, higher observed dust loadings can drive average losses as high as $2$ pp/day.
    
   Overall, the results suggest a relatively simple approach characterizing the statistical distributions of soiling losses using airborne dust measurements and short reflectance monitoring campaigns.

\end{abstract}




\end{frontmatter}


\section{Introduction}
    \input{sections/introduction.tex}

\section{Modeling} 
    \label{sec:modeling}
    \input{sections/modelling.tex}
\section{Experiments \& Results}
    \label{sec:results}
    \input{sections/results.tex}

\section{Discussion \& Conclusions}
    \label{sec:conclusions}
    Concentrating Solar Power (CSP) systems face challenges due to soiling losses, which must be accurately assessed at current and future plant sites. This study extends a previously developed physical soiling model by including 1) an uncertainty model for the deposition model, 2) a loss model based on Mie scattering and 3) Maximum Likelihood Estimation techniques to estimate model parameters from reflectance loss measurement and weather parameters. Two models were developed, one simplified and one physical, both of which were trained and tested on data sets obtained through experimental campaigns in different environments in Australia (Brisbane, Mount Isa, Wodonga).

    The results demonstrate the usefulness of the statistical perspective, with parameter estimates and their confidence intervals obtained via MLE methods, and prediction confidence intervals obtained that respond to changes in airborne dust. The simplified model achieved better generalization to other experimental campaigns and performs well on all three data sets. The physical model shows promise in fitting the deposition rate dependence on wind speed in the QUT experiments but performs significantly worse than the simplified model on the Mount Isa and Wodonga experiments.

    The results suggest that the simplified model can be used to obtain reasonable estimates of soiling rates by measuring airborne dust, undertaking a short reflectance monitoring campaign, fitting the simplified prediction model to the data, and using longer-term airborne dust measurements and the fitted model to predict the distribution of losses that would have been observed. The provided confidence intervals may be used to assess if more/longer experiments are needed for a particular site. Moreover, the results suggest that airborne dust measurements are important tools for understanding and predicting deposition rates, which has also been noted by other researchers \cite{micheli_investigation_2017}.
   
    There are some important limitations of this work. Firstly, there is some risk in ignoring the wind effect, which may be significant under different conditions at different sites. Indeed, the results and experimental data show that wind effects can be influential and modelled well by the physical model given the right data (e.g. the QUT data). Yet, the physical model in its current form clearly does not model wind effects accurately (e.g. Wodonga, Mount Isa). This is probably due in part to the simple model of the atmospheric boundary layer (ABL) and the limited understanding of key conditions (e.g. atmospheric stability, wind gusts) provided by the weather station measurements. Secondly, other assumptions about the composition, shape, and size distribution of the airborne dust will affect the prediction accuracy, but the relative influence and appropriateness of these assumptions for the studied sites needs to be examined. Thirdly, the incidence angle model in \eqref{eq:incidence_angle_model} is an approximation and there is some evidence for different functional dependence on the incidence angles \cite{heimsath_effect_2019}. Regarding the statistical distributions in Section~\ref{sec:distribution_results}, the use of the scenarios should be replaced by sampling from long-term dust data, when available. This will complete the picture of the statistical distribution airborne dust loadings and, in turn, the statistical distribution of daily losses.
    
    Future work will focus on developing a better understanding of the ABL and the nature of airborne dust using experimental measurements to identify and address the deficiencies of the physical model. Wet deposition and removal phenomena --- which may be dominant in humid conditions when airborne dust has significant solubility --- will also be explored. Moreover, experiments to assess the quality of the reflectance loss model will be pursued, and will be used to assess the need for a more sophisticated model. Finally, more experimental campaigns will be carried out across Australia and North America to expand the available data and a model benchmarking study will be carried out between the physical, simplified, and statistical approaches to identify their strengths and weaknesses.

\section{Acknowledgements}
    G. Picotti, M.E. Cholette, C.B. Anderson, and T.A. Steinberg acknowledge the support of the Australian Government for this study, through the Australian Renewable Energy Agency (ARENA) within the framework of the Australian Solar Thermal Research Institute (ASTRI – Project ID P54). The same authors also acknowledge the support of the U.S. Department of Energy's Solar Energy Technologies Office via the Soiling Subtask of the Heliostat Consortium (HelioCon). The authors would like to also acknowledge the support Kurt Drewes and Bruce Leslie of Vast Solar, and Paul Matuschka from Mars Petcare.

\appendix
\section{Fitting Quality Assessment}
\label{appendix_A}
\begin{figure*}[htp]
    \centering
    \begin{subfigure}[b]{0.45\textwidth}
        \includegraphics[width=\textwidth,keepaspectratio]{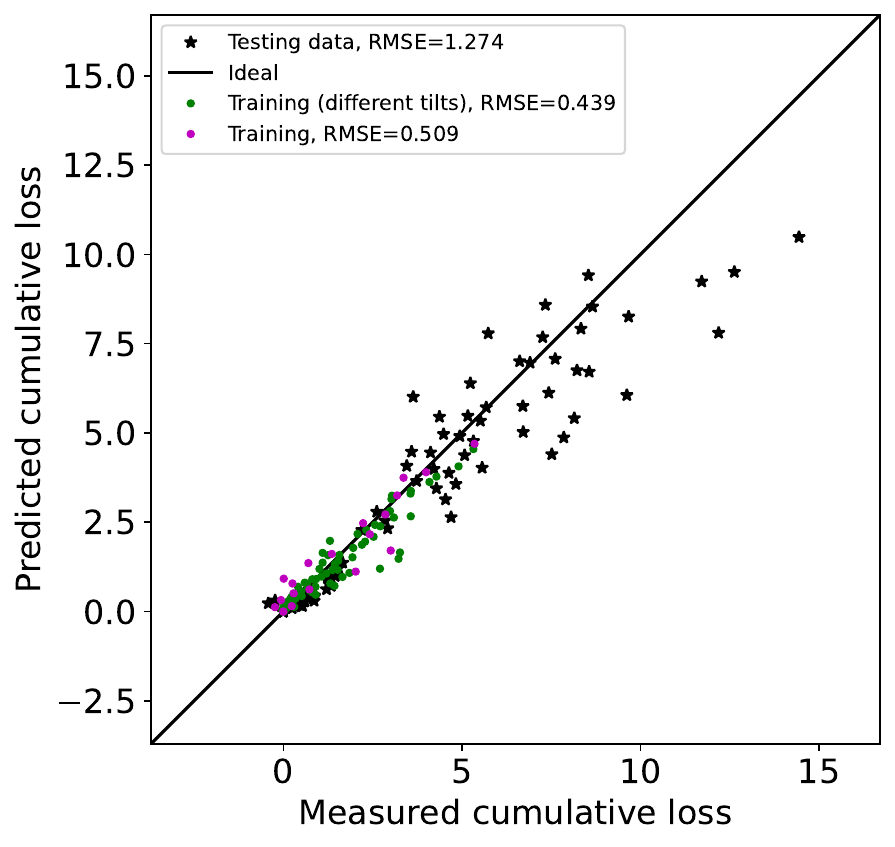}
        \caption{QUT}
    \end{subfigure}
    \hfill
    \begin{subfigure}[b]{0.43\textwidth}
        \includegraphics[width=\textwidth,keepaspectratio]{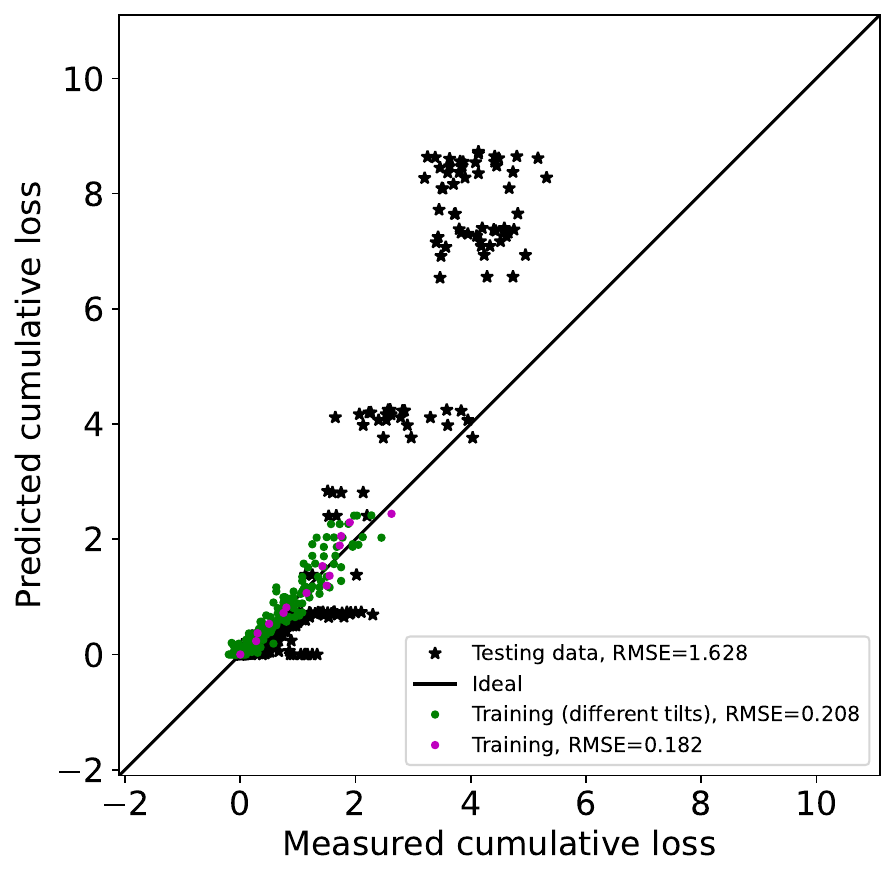}
        \caption{Mount Isa}
    \end{subfigure}
    \vspace{2ex}
    
    \begin{subfigure}[b]{0.43\textwidth}
        \includegraphics[width=\textwidth,keepaspectratio]{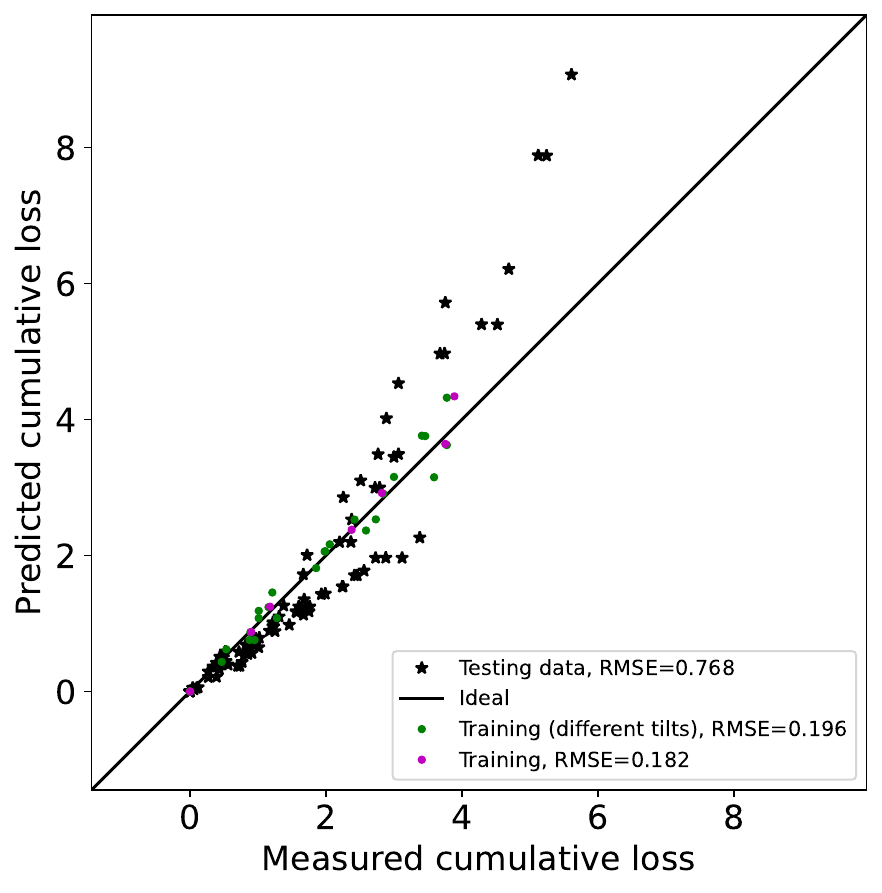}
        \caption{Wodonga}
    \end{subfigure}
    \caption{Cumulative reflectance loss predictions compared with reflectometer-measured losses for the \textbf{semi-physical} model. The black diagonal line denotes perfect prediction. Above (below) the line denote over (under) prediction. The root-mean-square error (RMSE) is in the legend.}
    \label{fig:fit_quality_sp}
\end{figure*}

\begin{figure*}[htp]
    \centering
    \begin{subfigure}[b]{0.45\textwidth}
        \includegraphics[width=\textwidth,keepaspectratio]{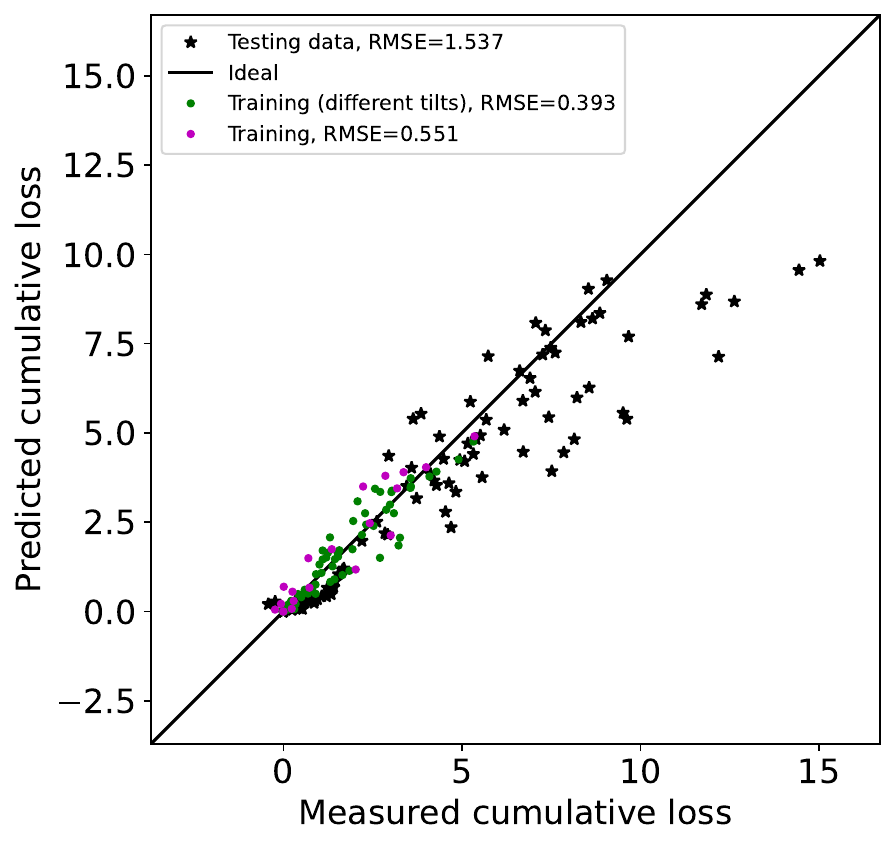}
        \caption{QUT}
    \end{subfigure}
    \hfill
    \begin{subfigure}[b]{0.425\textwidth}
        \includegraphics[width=\textwidth,keepaspectratio]{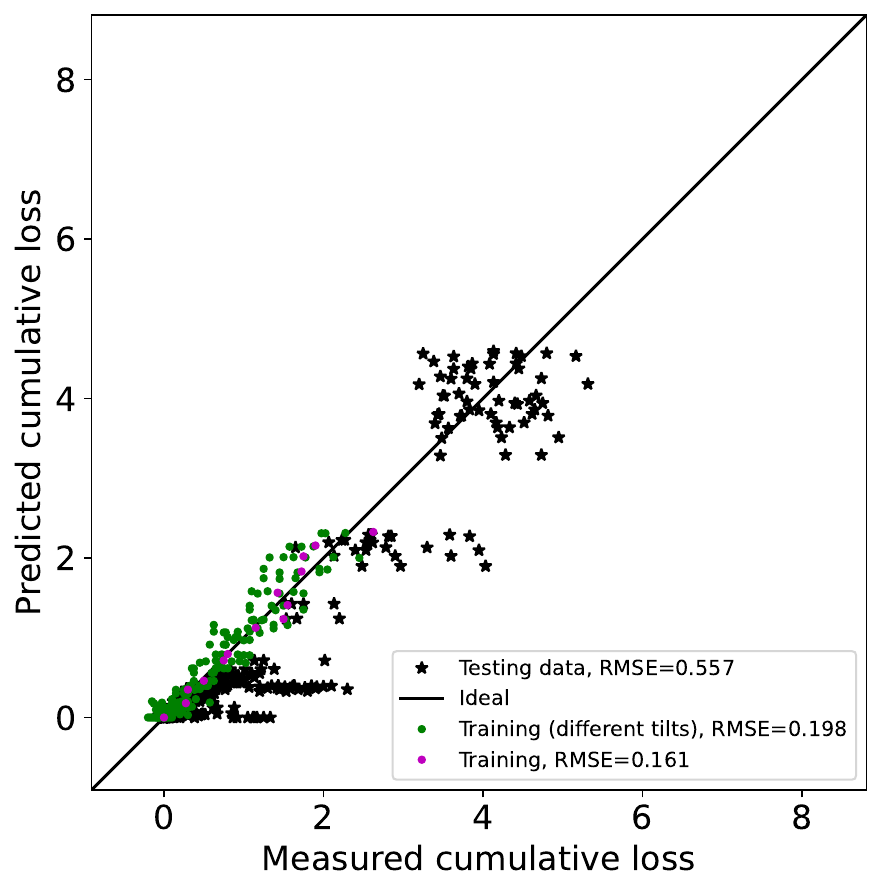}
        \caption{Mount Isa}
    \end{subfigure}
    \vspace{2ex}
    
    \begin{subfigure}[b]{0.425\textwidth}
        \includegraphics[width=\textwidth,keepaspectratio]{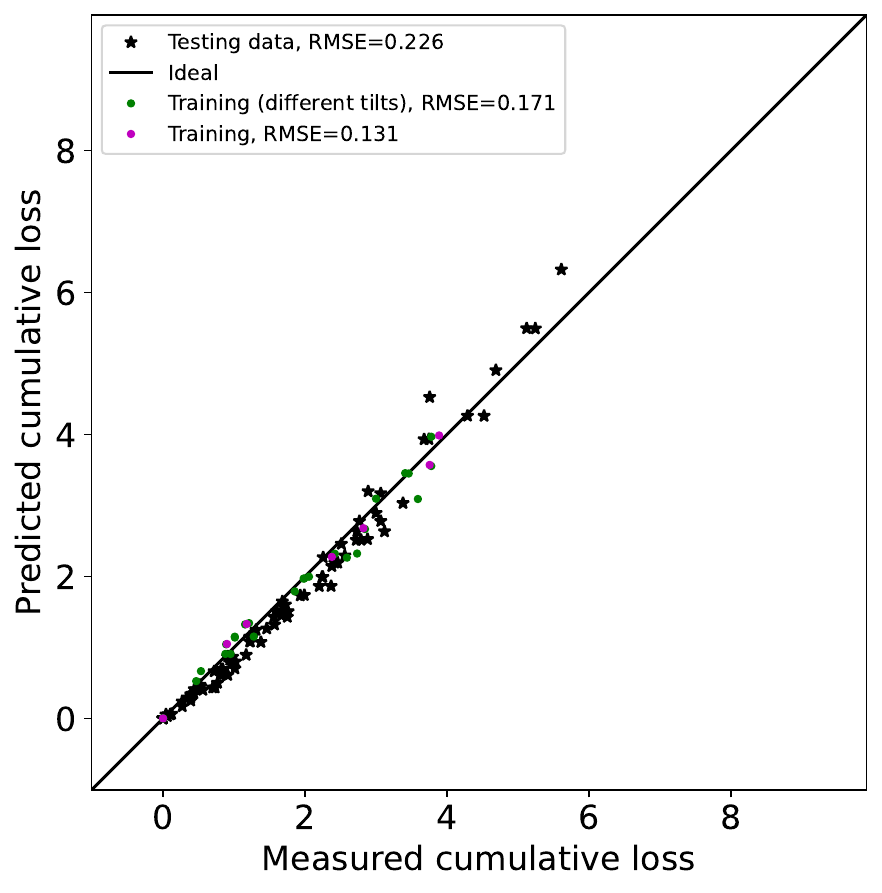}
        \caption{Wodonga}
    \end{subfigure}
    \caption{Cumulative reflectance loss predictions compared with reflectometer-measured losses for the \textbf{constant-mean} model. The black diagonal line denotes perfect prediction. Above (below) the line denote over (under) prediction. The root-mean-square error (RMSE) is in the legend.}
    \label{fig:fit_quality_cm}
\end{figure*}

In this section, some standard fit quality metrics and visualizations are presented to facilitate comparison with other studies. Figures \ref{fig:fit_quality_sp} and \ref{fig:fit_quality_cm} show the mean cumulative reflectance loss (predicted) plotted against reflectometer-measured losses\footnote{Reflectometer-measured losses are not precisely the "true" losses due to noise corruption} for the semi-physical and constant-mean simplified models, respectively. The magenta dots represent the data used for fitting, while the green dots represent data from different mirrors in the training interval. The black stars indicate data from all mirrors in the testing intervals. The root-mean-square error (RMSE) for each data set is shown in the legend of each plot. For all plots, the RMSE is smallest for the training data and largest for the testing data, reaching up to 1.6 percentage points for the testing data.
Regarding bias, the semi-physical model has predictions reasonably close to the ideal line for QUT and Wodonga. However, at higher measured losses, the data points are no longer symmetric around the ideal line, providing evidence that suggests the model under-predicts for QUT, and significantly over-predicts for Mount Isa and Wodonga when the losses are high. The over-prediction of losses is particularly evident for Mount Isa, with two large clusters corresponding to the high-dust event. This observation aligns with what was seen in Fig. \ref{fig:mount_isa_results_physical}, where this over-prediction resulted in a large drop in predicted reflectance. Another interesting feature in the Mount Isa portion of Fig. \ref{fig:fit_quality_sp} is the presence of two nearly horizontal lines of points near the bottom. These points correspond to the nearly-vertical mirrors, which observed significant measured losses despite the model predicting nearly zero losses for these tilts. 

The constant-mean model results in Fig. \ref{fig:fit_quality_cm} reinforce the observations in the main text: the QUT performance of the constant-mean model is similar to the semi-physical for the QUT experiments, but the constant-mean model performs better (in terms of RMSE and closeness to the ideal line) on the Mount Isa and Wodonga experiments. However, the two lines of points near the bottom of the Mount Isa plot remain, since the constant-mean model still neglects horizontal deposition.

\clearpage

\bibliographystyle{elsarticle-num} 
\bibliography{accepted_manuscript}

\end{document}

%% file: sections/introduction.tex
Maintaining high reflectance for Concentrating Solar Power (CSP) heliostats, is of paramount importance for the economics of the plant. One of the key degradation modes is the loss of reflectance due to the accumulation of dust on the surface of the heliostats. These losses are typically mitigated via artificial cleaning with brushes and (possibly high-pressure) water \cite{fernandez-garcia_equipment_2017}, which represent a significant Operation \& Maintenance (O\&M) cost for many plants \cite{mehos_concentrating_2020}. In an effort to optimally balance productivity losses against direct costs, a number of studies have been devoted to cleaning planning and resource allocation, including methods based on mathematical programming \cite{wales_optimizing_2021, picotti_optimization_2020} or Markov Decision Processes (MDPs) \cite{terhag_optimization_2019, truong-ba_sectorial_2020}. Yet, these approaches rely on estimation of the local soiling dynamics, which is influenced by local environmental factors, exhibits significant randomness and may be  affected by seasonality. As a result of these factors, local soiling rates for CSP reflectors may vary from a few tenths of a percentage point to a few percentage points per day \cite{alami_merrouni_csp_2020,picotti_soiling_2018,micheli_economics_2021} --- a variation that would require vastly different cleaning strategies and strongly influcence the economic viability of a plant. Therefore, accurately characterizing and predicting the dependence of these soiling losses based on location-specific context is essential to plan cleaning approaches and reduce the financial risk during site selection \cite{zhu_roadmap_2022}.

A number of soiling predictive models have been developed in literature for both PV and CSP \cite{picotti_soiling_2018,costa_solar_2018}. A main distinction can be made between ``physical" and ``statistical" approaches. The statistical approaches typically utilize extensive experimental data provided by reflectance measurements from hand-held reflectometers \cite{bonanos_characterization_2020} or custom semi-automated devices (e.g. the TraCS \cite{conceicao_csp_2018} or AVUS \cite{heimsath_automated_2018}) together with weather measurements (e.g. wind speed, humidity, PM10, etc.). A model is then constructed via regression analysis \cite{micheli_investigation_2017,bonanos_characterization_2020,javed_modeling_2017}, Artificial Neural Networks \cite{bonanos_characterization_2020,javed_modeling_2017}, autoregressive models \cite{ballestrin_soiling_2022}, linear discrete-time state-space models \cite{bouaddi_soiled_2015,bouaddi_comparative_2017}, Markov Regime Switching Models \cite{bouaddi_modeling_2018}, or other data-driven approach to describe the dependence of soiling losses on a number of environmental parameters. While these statistical approaches avoid the need for a detailed model, they do not provide any insight on the relevant physical processes and extrapolation to novel weather conditions or at novel sites is dubious. 

On the other hand, physical models describe the deposition of dust particles and their interaction with solar collectors surfaces through physical laws that describe the individual mechanisms that happen in the different stages of the overall process. Picotti et al. \cite{picotti_soiling_2018,picotti_development_2018} developed and validated a model for CSP soiling losses which divided the process into four main phases (dry deposition, adhesion, removal, and reflectance losses). Each was separately modelled through physical and/or geometrical equations. Wolfertstetter et al. \cite{wolfertstetter_modelling_2019} adopted a similar deposition model and applied this model on two data sets. A prediction root-mean-square error (RMSE) of 0.44 \%/day was estimated using cross validation, but a detailed statistical characterization of the reflectance loss predictions was not conducted. Lozano-Santamaria et al. \cite{lozano-santamaria_stochastic_2020} applied similar considerations to assess the dust deposition on air coolers in CSP (fouling), demonstrating the relevance of many of those environmental parameters on the deposition process. Other authors focused on the impact of deposited dust on reflectance losses, exploiting the concept of ``turbid medium" and the Beer-Lambert law to identify the impact of different incidence angles \cite{heimsath_effect_2019}, or applying the Mie scattering theory to measured gravimetric density and assumed deposited dust size distribution to compute the cleanliness of the solar collectors (both PV and CSP) \cite{bellmann_comparative_2020}. 

In PV literature, physical models have also been employed in a number of studies. You et al. \cite{you_temporal_2018} computed the dry deposition velocity to assess airborne dust deposition and subsequently applied a linear relation between deposited dust and efficiency loss. Following a very similar approach, Coello and Boyle \cite{coello_simple_2019} computed the dry deposition velocity using standard models from atmospheric literature, and eventually assessed the soiling ratio as a function of total mass accumulated on the PV panels. Fernandez-Solas et al. \cite{fernandez-solas_estimation_2022} managed to exploit the physics of PV modules to assess soiling form their electrical available measurements. These latter approaches grounded in physical models have the advantage that they offer insight into the relevant physical processes, but the simplified models typically used for key phenomena (e.g. a logarithmic profile for wind speeds, atmospheric stability) are difficult to validate and still require some experimental data from the site for tuning to produce accurate predictions.  

Taken together, the existing studies have demonstrated the promise of both statistical and physical approaches to predict soiling losses, with the former offering accurate predictions and the latter offering more physical insight. However, there are a number of gaps:
\begin{enumerate}
    \item \textbf{Statistical characterization of soiling losses}. While many of the existing models assess the quality of the fit (e.g. via mean-square error of forecasts), there has been little effort on modelling the impact of the various sources of uncertainty on the resulting soiling rate predictions (e.g. provide prediction confidence intervals), despite the significant impact that such uncertainty has on cleaning operations \cite{truong-ba_sectorial_2020}. This is particularly true for physical soiling models \cite{picotti_development_2018,wolfertstetter_modelling_2019} which do not include error terms, but is also true of the statistical models, which tend to focus on forecasts using expected values rather than uncertainties. A more complete stochastic description of different uncertainties in the soiling process and measurements can yield 1) insights into how these uncertainties affect the parameter estimates (i.e. parameter confidence intervals), and 2) statistical confidence/credible intervals on predictions to quantify soiling loss risks. 
    \item \textbf{Use of short-term reflectance measurement campaigns}. For existing CSP studies, while the predictions are of good quality, they have only been evaluated at a small number of sites (typically one or two) using long-term data sets ($\sim$4-48 months) \cite{wolfertstetter_modelling_2019, bouaddi_comparative_2017,bouaddi_modeling_2018,conceicao_csp_2018}. Such long-term reflectance campaigns either require laborious measurements from hand-held reflectometers \cite{picotti_development_2018,bonanos_characterization_2020,bouaddi_modeling_2018} or elaborate custom devices \cite{wolfertstetter_novel_2012}. It is therefore advantageous to explore the possibility of using shorter-term reflectance campaigns to assess the soiling rates at a particular site. Such short term experiments would clearly have a lower accuracy, but by using the aforementioned statistical analysis, the accuracy of the predictions could be estimated and inform a decision on whether or not more data needs to be collected.   
    \item \textbf{Understanding soiling losses in Australia}. Existing CSP soiling loss studies have been primarily conducted in Europe and North Africa, and soiling studies from other CSP-relevant locations (e.g. Australia) are lacking.  
    \item \textbf{CSP reflectance loss models are simplistic}. Existing physical CSP soiling models use geometrical area coverage to assess the reflectance loss of the deposited dust \cite{picotti_development_2018,wolfertstetter_modelling_2019}. Yet, it is well known that the losses are governed by Mie Scattering \cite{bellmann_comparative_2020,roth_effect_1980}, and thus the scattering effective area is likely significantly different than the aerodynamic cross-section. It is therefore advantageous to integrate the Mie Scattering models (e.g. similar to \cite{roth_effect_1980,bellmann_comparative_2020} with the physical deposition models. 
\end{enumerate}

In this paper, two new stochastic soiling loss models for solar mirrors --- a \emph{semi-physical} and simplified \emph{constant-mean} model --- are developed that aim at a characterization of uncertainty in the soiling process by extending the physical model previously developed in \cite{picotti_development_2018}. The contributions are as follows: 1) a novel uncertainty models for the deposition velocity and measurement errors, 2) improvement of the CSP reflectance loss model to include Mie scattering, 3) derivation of expressions for the statistical distribution of soiling losses and Maximum Likelihood Estimation (MLE) methods, and 4) a novel simplified model based on the insights from the first three developments. These developments enable, for the first time, the combination of physical insight with an assessment of parameter estimation uncertainties induced by the uncertain deposition and measurements. Moreover, both the semi-physical model and the simplified model are deployed on a total of ten experimental campaigns taken at three different sites around Australia with significantly different climates. The results permit an evaluation of the suitability of the models to describe CSP-relevant soiling losses across different sites using short ($\sim$1 week) soiling campaigns and longer-term, automatically acquired weather data. 

The remainder of the paper is organized as follows. Section \ref{sec:modeling} describes the physical soiling model, which is subsequently extended to describe the statistical distribution of reflectance changes and develop a simplified model. Section \ref{sec:results} describes the ten experimental campaigns as well as the parameter estimation and testing of the developed models. Finally, Section \ref{sec:conclusions} concludes the paper and suggests areas for future work.

%% file: sections/modelling.tex
The models developed in this paper are comprised of two components: a deterministic physics-based deposition model based on an extension of the authors' earlier work \cite{picotti_development_2018}, and a wholly novel stochastic component. Each component will be discussed in the sequel, and an expression for the statistical distribution of reflectance changes will be developed. This development will serve to 1) clarify some important phenomena for the observed reflectance of CSP collectors subject to soiling and 2) enable the development of the statistical likelihood of observations, which is the basis for the estimation of the model parameters. 

\subsection{Physical model}
    Firstly, the physical model will be discussed. While the deposition model largely follows \cite{picotti_development_2018}, the discussion here will focus on details that are pertinent to later developments and will introduce a refined loss model based on Mie Theory, following approaches similar to \cite{bellmann_comparative_2020,roth_effect_1980}.
    \subsubsection{Reflectance Loss Model}
        Consider the following model for the reflectance of the heliostat\footnote{The wavelength dependence is omitted here. The surface is assumed to be grey or the reflectance is assumed to be a solar weighted average.}: 
            \begin{align}
                \rho\left(A_{soil}(t),\phi \right) = \rho_0 \left[ 1-\frac{A_{soil}(t)}{A_{ref}} \cdot h(\phi) \right]
                \label{eq:reflectance}
            \end{align}
        where $A_{soil}$ is the effective soiled area when the beam incidence is normal to the surface (i.e. $\phi=0$) \footnote{As is typical, particles are assumed to be spherical.}, $\phi$ is the angle of incidence at which the reflectance is assessed/measured, $\rho_0$ is the nominal reflectance (assumed to be directionally uniform), $A_{ref}$ is the unsoiled reflective area\footnote{Without loss of generality, it is assumed that $A_{ref}=1$ in this study.}, and $h(\phi)\geq 1$ is an incidence factor that accounts for the additional blocking and shading when $\phi>0$. This factor depends on the nature of the reflector, particularly whether it is a first- or second-surface reflector --- a fact which will be discussed later. The separation of the soiling and incidence angle effects ignores the potential mutual blocking and shading of an area of the reflective surface. As a consequence, \eqref{eq:reflectance} is likely valid for lightly soiled mirrors at sufficiently low incidence angles. These caveats are likely acceptable for ST plants, where heavy soiling is uncommon due to regular cleaning.
        
        In the absence of cleaning (and rain)\footnote{Cleaning/rain models are considered outside the scope of this paper.}, the total soiled area of the deposited particles can be computed as:
            \begin{align}
                A_{soil}(t) = A_{soil}(t_0) + \frac{\pi}{4} \int_{0}^{\infty}{D^2 N(t_0,t,D) \gamma(D,\phi_a) dD}
                \label{eq:A_soil}
            \end{align}
        where $D$ is the particle diameter and $N\left(t_0,t,D \right)$ is the number of particles of diameter $D$ accumulated on the mirror between $t_0$ and $t$, $\phi_a$ is the acceptance half-angle of the reflectance measurements (which is assumed to be constant in this study). The function $\gamma(D,\phi_a)$ is a weighting factor that is calculated based on Mie Theory \cite{roth_effect_1980}
            \begin{align}
            \gamma(D,\phi_a) \triangleq 2\pi \int_0^\infty Q_{ext}\left(\frac{\pi D}{\lambda},m\right)\, g\left(\phi_a\right) \bar{E}(\lambda) d\lambda
        \end{align}
        where  $m$ is the complex refractive index of the dust,  $Q_{ext}\left(\frac{\pi D}{\lambda},m\right)$ is the typical extinction coefficient from Mie Theory, which is the ratio of the extinction area and the cross-sectional area of the particle. Finally,
        \begin{align}
                g\left(\phi_a\right)  & \triangleq \int_{-1}^{\cos(\phi_a)} p\left(\mu\right)  d\mu
        \end{align}
        where $p\left(\mu\right)  $ is the scattering function\footnote{The scattering function used here is normalized to integrate to one.}. This last integral accounts for the small, but finite, acceptance angle of the measurement device.
    
        This area loss model is similar to \cite{picotti_development_2018} with one important extension: the weighting factor $\gamma(D,\phi_a)$ allows particles of different diameters to have different effective cross-sectional areas for light scattering --- a fact that may be important for the medium-sized particles (0.1-10$\mu m$) which have dimensions similar to the wavelengths of the incident light. Indeed, this model simplifies to that of \cite{picotti_development_2018} if $\gamma(D,\phi_a)=1$ for all  $D$.  

        When the reflectance measurement angle is normal ($\phi=0^\circ$), the reflectance lost is simply the effective cross-sectional area, i.e. $h(\phi)=1$ in \eqref{eq:reflectance}. However, for non-zero incidence, $h(\phi)\neq1$ and an expression must be developed. For this work, a simplified geometrical model (albeit using the effective cross-sectional area from the Mie calculations) is used to model the incidence angle effect. In the case where the dust is deposited directly on the reflector surface, $h(\phi) = \frac{1+\sin(\phi)}{\cos(\phi)}$ \cite{picotti_development_2018}. For a second-surface reflector with a covering glass a few millimeters thick with incidence angles sufficiently far from zero (e.g. greater than $1^{\circ}$), the total area loss is well-approximated by
        \begin{align}
  		    h(\phi) = \frac{2}{\cos(\phi)} \label{eq:incidence_angle_model}
        \end{align}
        where the two in the numerator is to account for the blocking and shading of the reflective surface \cite{heimsath_effect_2019,roth_effect_1980}. Therefore, once the number of particles at each diameter are known at some $t_0$, the reflectance can be computed using \eqref{eq:reflectance}. It is likely that $t_0$ will be just after cleaning and $A_{soil}(t_0)=0$ if perfect cleaning is assumed.
        
    \subsubsection{Deposition Model}
        Due to the measurements available for airborne dust, it is convenient to express the number distribution as a function of the cumulative \emph{mass} deposited on the mirror $M(t_0,t,D)$
            \begin{align}
                N\left(t_0,t,D\right) = \frac{M(t_0,t,D)}{\frac{1}{6}\rho\pi D^3}.
                \label{eq:number_as_a_function_of_mass}
            \end{align}
        which assumes a constant density. All particles are assumed to Quartz spheres, since it is the main component of dust worldwide and most other major components (Calcite, Dolomite, Kaolinite) have similar densities and refractive indices \cite{holsen_dry_1992,ilse_fundamentals_2018}. The mass deposited is modeled as
            \begin{align}
                M(t_0,t,D) = \int_{t_0}^{t} f(\tau,D) d\tau
                \label{eq:cumulative_mass}
            \end{align}
        where $f(\tau,D)$ is the particle mass flux for particles of diameter $D$ at time $\tau$. The mass flux is modelled as product of the mass of the particles in the air, the vertical velocity of the particles, and the mirror horizontal projected area \cite{picotti_development_2018}: 
            \begin{align}
        		f(t,D) = m(t,D)\cdot v_{d}(w(t),T(t),D;\,hrz0)\cdot \cos(\theta(t)) 
                \label{eq:flux}
        	\end{align}
    	where $m(t,D)$ is the mass concentration distribution of airborne particles of diameter $D$, $\theta(t)$ is the tilt angle of the mirror, and $v_d(w(t),T(t),D;\,hrz0)$ is the deposition velocity , which include gravitational settling, inertial effects, and Brownian motion, depending on particles diameter, as described in \cite{picotti_development_2018}. This last term is based on the well-known resistance model \cite{seinfeld_atmospheric_1998} and has a functional dependence on the diameter, wind speed $w(t)$, air temperature $T(t)$, and the site-dependent parameter $hrz0$ (the ratio of a reference height to the surface roughness of the site). Beyond this functional dependence, the details of the model are not particularly important for the later developments in this paper, so the interested reader is referred to \cite{picotti_development_2018} for more details.
        
        Unfortunately, the mass distribution in \eqref{eq:flux} is not usually known. Typically, the cumulative mass up to a few diameters --- often just one --- are measured. The most commonly available are $PM_{10}$ and $TSP$, which are the cumulative particle mass up to a diameter of 10$\mu$m, and the total suspended particles, respectively. Assuming constant density, the mass distribution is factored into the following form
            \begin{align}
            	m(t,D) = \rho \frac{1}{6}\pi D^3 \cdot \alpha(t) \cdot \hat{n}(D)
            \label{eq:mass_distribution}
            \end{align}
        where $\hat{n}(D)$ is the well-known tri-modal log-normal number distribution \cite{Seinfeld2016}\footnote{The $\ln(10)\,D$ term is due the transformation of the independent variable from $\log D$ to $D$. The interested reader is referred to \cite[p. 333]{Seinfeld2016} for details.}:
            \begin{align}
            \hat{n}(D) = \sum_{i=1}^3 \frac{N_i}{\ln(10)\,D\,\sqrt{2\pi}\log{\sigma_i}}\exp\left[ -\frac{(\log(D)-\log(\mu_i))^2}{2 (\log(\sigma_i))^2} \right]
            \label{eq:prototype_number}
            \end{align}
        where $N_i$, $\mu_i$, and $\sigma_i$ are parameters of the different modes. Prototype parameters can be selected based on the nature of the site (e.g. urban, rural, etc.) from literature \cite{seinfeld_atmospheric_1998}. The time-dependent term $\alpha(t)$ is the ratio of the relevant airborne dust measurement with its corresponding value for the prototype mass distribution. To be more specific, let $c_\delta(t)$ be the concentration of airborne dust with diameters less than $\delta$, then $\alpha(t)$ can be expressed as
            \begin{align}
                \alpha(t) = \frac{c_\delta(t)}{\displaystyle \frac{\pi\rho}{6} \int_{0}^{\delta} D^3 \cdot \hat{n}(D) \, dD}.
                \label{eq:alpha_scale}
            \end{align}
        In other words, $\alpha(t)$ is the ratio of an observed mass concentration below diameter $\delta$ and the corresponding value of the mass concentration implied by the prototype distribution. Thus, the prototype distribution is ``scaled'' according to the measurements. 
        
        Putting \eqref{eq:number_as_a_function_of_mass}---\eqref{eq:alpha_scale} together yields:
            \begin{align}
                N\left(t_0,t,D\right) = \hat{n}(D) \cdot  \int_{t_0}^{t} \alpha(\tau) \cdot v_{d}(w(\tau),T(\tau),D;\,hrz0)\cdot \cos(\theta(\tau))    \,  d\tau
                \label{eq:cumulative_number_integral}
            \end{align}
        This integral may be computed approximately by discretizing the timeline into $\Delta t$ intervals. Let interval $k$ denote the time interval $[k\Delta t,(k+1)\Delta t)$ and assume that the measurements are taken at the beginning of the interval. Moreover, assume that the time-varying quantities are constant at their average values for the interval. The integral \eqref{eq:cumulative_number_integral} can be approximated as:
            \begin{align}
                N(k_0\Delta t,k\Delta t,D) \approx \hat{n}(D) \cdot  \sum_{i=k_0}^{k-1} \alpha_i \cdot v_{d}(w_i,T_i,D; hrz0)\cdot \cos(\theta_i) \Delta t
                \label{eq:cumulative_number_sum}
            \end{align}
        where $w_i\triangleq w(i\Delta t)$ and an analogous definition holds for each of $\alpha_i$, $w_i$, $T_i$ , and $\theta_i$.
        
       Equation \eqref{eq:cumulative_number_sum} enables the computation of the number of particles that fall on a surface between two times. However, it is possible that these particles are removed from the surface at some point due to wind, sliding, rolling, or other removal forces \cite{ilse_fundamentals_2018}. For the purposes of this model, only competition between the rolling moment and gravity removal is considered --- resulting in a critical diameter $D_c(\theta)$ above which all particles are removed \cite{picotti_development_2018}. For this study, it is assumed that there is a $D_c^\prime$ above which any deposited dust is removed via rolling either instantaneously (for a fixed tilt mirror) or within a day (for a tracking mirror). Such critical diameter can be computed for any dust particle size and any tilt angle as described in  \cite{picotti_development_2018}. This is a reasonable assumption in two important situations:
            \begin{enumerate}
                \item Fixed tilt mirrors (e.g. for soiling experiments), where $\theta=constant$.
                \item When tracking mirrors frequently achieve some high tilt angle. This can happen if mirrors are either stowed at a high tilt (close to $90^\circ$) or achieve some high tilt as a part of the stowing process (e.g. they are near vertical during the manoeuvre to the face-down stowing position).
            \end{enumerate}
        In both of these cases,  dust that is regularly removed is simply ignored and the dust accumulated on the mirror is
            \begin{align}
            	N(k_0\Delta t,k \Delta t,D) \approx 
            	\begin{cases}
                    \text{Eq. \eqref{eq:cumulative_number_sum} } & D\leq D_c^\prime \\ 
                    0 & D > D_c^\prime
                \end{cases}
                \label{eq:cumulative_deposited_below_critical_diameter}
            \end{align}
            
\subsection{Stochastic model \& the distribution of changes in measured reflectance}
    The physical model is capable of utilizing weather and airborne dust measurements and a few key choices (prototype distribution, $hrz0$) to  provide predictions of reflectance losses at different angles of incidence. However, these \emph{deterministic} predictions are subject to a number of assumptions (e.g. spherical particles, constant density) and thus the deposited dust is likely subject to significant uncertainty. Moreover, the reflectance measurements (particularly for soiled samples) are inherently uncertain. In the remainder of this section, a model will be developed to attempt to quantify the effects of uncertainty in the deposition flux and measurements.
    
    Consider the reflectance measurements at time index $k$, $r_k$. Because the reflectometer-measured average reflectance has significant uncertainty (due to both the device repeatability and the random distribution of dust particles on the reflective surfaces), the measured values are modelled as
    \begin{align}
        r_k = \rho\left(A_{soil,k},\phi_k \right) + \epsilon_{r,k}
    \end{align} 
    where $\epsilon_{r,k} \sim \mathcal{N}\left(0,\sigma_{r,k}^2 \right)$ is allowed to vary with time and $k \subseteq \left\{0,1,\ldots \right\}$ are the time indices where the reflectance is measured. The measurement noise variances $\sigma_{r,k}^2$ can be estimated by taking repeated measurements at different locations and computing the estimated variance of the mean. The time index is included since this variance has been noted to change as the mirror becomes more soiled. 
    
    Now, consider two consecutive measurement taken at time indices $\ell<m$. The change in the reflectance loss is:
        \begin{align}
        r_{m} - r_{\ell}      & = \rho\left(A_{soil,m},\phi_m\right)-\rho\left(A_{soil,\ell},\phi_{\ell}\right) + \epsilon_m - \epsilon_\ell \nonumber \\
                            & = \frac{\rho_0}{A_{mirror}}\left[ h\left(\phi_\ell\right) A_{soil,\ell} - h\left(\phi_m\right) A_{soil,m} \right]+ \epsilon_{r,m} - \epsilon_{r,\ell}
                            \label{eq:reflectance_difference}
        \end{align}
    Let $g(w_j,T_j,D;\,hrz0) \triangleq \alpha_j\, \hat{n}(D)\, v_d(w_j,T_j,D;\,hrz0)$ and assume that the soiled area at $t_0=k_0\Delta t$ is known\footnote{e.g. the mirror is cleaned perfectly at $t_0$ and $A_{soil,k_0}=0$.}. The soiled area at any $k>k_0$ can be computed as:
        \begin{align}
            A_{soil,k}  & = A_{soil,k_0} + \frac{\pi}{4} \int_{0}^{\infty} D^2\,N(k_0\Delta t,k\Delta t,D) \gamma(D,\phi_a) dD \nonumber \\
                        & = A_{soil,k_0} + \frac{\pi}{4} \Delta t  \sum_{i=k_0}^{k-1} \cos(\theta_i) \, \int_{0}^{D_{c}} D^2\,\gamma(D,\phi_a)\,g(w_i,T_i,D;\,hrz0) dD  \nonumber \hfill \\
                        & = A_{soil,k_0} + \sum_{i=k_0}^{k-1} \alpha_i \cdot \mu(w_i,T_i;\,hrz0) \cdot \cos(\theta_i)
        \end{align}
    where $\mu(w_i,T_i;hrz0) \triangleq \frac{\pi}{4} \Delta t \int_{0}^{D_{c}} D^2\,\gamma(D,\phi_a)\,\hat{n}(D) \cdot v_d(w_i,T_i,D;hrz0) dD$. Since the area loss rate has significant uncertainty, it is reasonable to include a noise term in the following manner:
        \begin{align}
            A_{soil,k} = A_{soil,k_0} + \sum_{i=k_0}^{k-1} \alpha_i \cdot \cos(\theta_i) \cdot \left[ \mu(w_i,T_i;hrz0) + \varepsilon_{i} \right] \label{eq:stochastic_cumulative_soiled_area}
        \end{align}
    where $\varepsilon_i \sim \mathcal{N}(0,\sigma_{dep}^2)$ are assumed to be i.i.d. noise terms. In other words, the model captures the mean of the reflective area loss, while the variance in this loss is modelled via $\sigma_{dep}^2$. The distribution of $A_{soil,k}$ can now be easily established as the sum of linearly transformed random variables 
        \begin{align}
            A_{soil,k} \sim \mathcal{N}\left(m_k,s_k\right)
        \end{align}
    where
        \begin{align*}
            m_k & \triangleq A_{soil,k_0} + \sum_{i=k_0}^{k-1} \alpha_i \, \cos(\theta_i) \, \mu(w_i,T_i;hrz0) \\
            s^2_k & \triangleq \sigma_{dep}^2 \sum_{i=k_0}^{k-1} \alpha_i^2 \, \cos^2(\theta_i)
        \end{align*}
    We can now compute the change in reflectance using only observable quantities using \eqref{eq:reflectance_difference}:
        \begin{align}
            r_m - r_{\ell}  & = b(\phi_\ell) A_{soil,\ell} - b(\phi_m) A_{soil,m} + \epsilon_m - \epsilon_\ell \nonumber \\ 
                            & = \left[ b(\phi_\ell)-b(\phi_m)\right] A_{soil,\ell} \nonumber \\
                            & \qquad - b(\phi_m) \sum_{j=\ell}^{m-1} \alpha_j\,\cos(\theta_j) \cdot (\mu(w_j,T_j; hrz0) + \varepsilon_{j}) +\epsilon_m - \epsilon_\ell \label{eq:reflectance_change}
        \end{align}
    where $b(\phi) = \frac{\rho_0}{A_{mirror}}h(\phi) $. Equation \eqref{eq:reflectance_change} is interesting in its own right. The first term describes how a change in measurement angle will yield a change in observed reflectance, even when no additional dust is deposited. The second term accounts for the losses due to the additional dust deposited from time index $\ell$ to index $m$, and the last two terms model the imperfect measurements. In the case of a tracking heliostat, the incidence angle will change significantly throughout the day and this first term will significantly contribute to the observed changes in reflectance. 
    
    However, reflectance is typically measured using a fixed-incidence-angle reflectometer, and so $\phi_\ell = \phi_m$ and thus $\left[ b(\phi_\ell)-b(\phi_m)\right]=0$, so the first term has no effect on the observed reflectance change. The fixed-incidence condition --- together with the fact that \eqref{eq:reflectance_change} is an affine transformation of normal random variables --- implies that the statistical distribution of the reflectance changes may be written as:
        \begin{align}
            r_m - r_{\ell} & \sim \mathcal{N}\left( \mu_{m,\ell}, \sigma_{m,\ell}^2 \right) \label{eq:difference_distribution}
        \end{align}
    where
        \begin{align}
            \mu_{m,\ell}        & = -b(\phi_m)\sum_{j=\ell}^{m-1} \alpha_j\,\cos(\theta_j) \cdot \, \mu(v_j,T_j;hrz0) \label{eq:loss_mean} \\[2ex]
            \sigma_{m,\ell}^2   & = \sigma_{dep}^2 \, b(\phi_m)^2\, \sum_{j=\ell}^{m-1} \alpha_j^2 \,\cos^2(\theta_j) + \sigma_{r,m}^2+\sigma_{r,\ell}^2 \label{eq:loss_variance}
        \end{align} 
    This expression permits the evaluation of the probability density of reflectance changes given meteorological parameters (airborne dust concentration, the ambient temperature, wind speed), measurement parameters (acceptance and incidence angles), and the tilt history.
\subsection{Likelihood \& parameter estimation}

  Consider the data set $\mathcal{D}$, which consists of data from $L$ experiments each consisting of measurements taken at time indices $k_\ell, \, \ell =  1,2,\ldots,N_\ell$ where $k_j > k_i \; \forall j>i$. Let $\bar{w}_i = \left(w_i,T_i,c_i,\theta_i\right)$ be the tuple of input variables at time index $i$ and let $hrz0=h$ and $\sigma_{dep}^2 = q$ . The data can be written as:
    \begin{align}
        & \mathcal{D}     = \bigcup_{\ell=1}^L \mathcal{D}_\ell \\
        & \mathcal{D}_\ell   = \left\{ \left(\Delta r_{\ell,k_{i+1}} ,\bar{w}_{\ell,k_i},\bar{w}_{\ell,k_i+1},\ldots,\bar{w}_{\ell,k_{i+1}-1},\bar{w}_{\ell,k_{i+1}}\right),\, i=0,1,\ldots,N_\ell-1  \right\}
    \end{align}
    where $\Delta r_{\ell,k_i} = r_{\ell,k_i} - r_{\ell,k_{i-1}}$ is the reflectance loss of experiment $\ell$ between successive measurements. The likelihood for this data is
    \begin{align}
        L\left( \mathcal{D} \mid h,q \right)    & = \prod_{\ell=1}^{L} \prod_{i=1}^{N_\ell} p(\Delta r_{\ell,k_i}\mid h,q)
    \end{align}
    where $p(\cdot\mid h,q)$ is the density for the normal distribution from \eqref{eq:difference_distribution} with $hrz0=h$ and $\sigma_{dep}=q$. Using this probability model, the log-likelihood can be written as
    \begin{align}
        \ell\left( \mathcal{D} \mid h,q \right) = \sum_{\ell=1}^L \sum_{i=1}^{N_\ell}-\frac{1}{2}\log(2\pi) - \frac{1}{2}\log\left(\sigma_{\ell,k_i,k_{i-1}}^2\right) - \frac{(\Delta r_{\ell,k_i} -\mu_{\ell,k_i,k_{i-1}},)^2}{2\sigma_{\ell,k_i,k_{i-1}}^2} \label{eq:likelihood}
    \end{align}
    where  $\mu_{\ell,k_i,k_{i-1}}$ and $\sigma_{\ell,k_i,k_{i-1}}^2$ are the mean and variance of the reflectance loss for the $i$th difference in the $\ell$th experiment and which can be computed from \eqref{eq:loss_mean} and \eqref{eq:loss_variance}. Note that neither $h$ nor $q$ appear explicitly here, but they are embedded in the computation of $\mu_{\ell,k_i,k_{i-1}}$ and $\sigma_{\ell,k_i,k_{i-1}}^2$ in \eqref{eq:loss_mean} and \eqref{eq:loss_variance}.
    
    The following procedure can be used to compute the likelihood:
    \begin{enumerate}
        \item Compute $D_c$ for the fixed tilt or stow angle. 
        \item Compute $\mu(w_i,T_i)$ with $hrz0=h$, setting and $v_{d}(w_i,T_i,D^\prime;\,h) = 0$  $\forall D^\prime\geq D_c$.
        \item \label{step:compute_likelihood} Compute $\mu_{\ell,k_i,k_{i-1}}$ via \eqref{eq:loss_mean} and $\sigma_{\ell,k_i,k_{i-1}}^2$ via \eqref{eq:loss_variance} with $\sigma_{dep} = q$ for all $i=1,2\ldots,N_\ell$ and sum.
        \item Repeat \ref{step:compute_likelihood} for $\ell=1,2,\ldots,L$ and sum result.
    \end{enumerate}

    For estimation, the measurement noise variances $\sigma_{r,k}^2$ can be estimated separately using repeated measurement on the same reflector and the sample variance of the mean. The two remaining free parameters --- $hrz0$ and $\sigma_{dep}$ --- were estimated via maximization of \eqref{eq:likelihood}:
        \begin{align*}
            \widehat{hrz0},\hat{\sigma}_{dep} & = \arg\max_{h,q} \ell\left( \mathcal{D} \mid h,q \right)
        \end{align*}
    Because of the bounds on the variables, the optimization decision variables were $\log\left[ \log\left( hrz0 \right) \right]$ and $\log\left( \sigma_{dep} \right)$ to ensure that the parameters obey the physical limits of $hrz0>1$ and $\sigma_{dep}>0$ during optimization. The parameter confidence intervals are found using Fisher Information in the log-transformed space and inverting the transformation for the upper and lower bounds. 

\subsection{Constant mean deposition velocity simplification}
    The above development suggests the following simplification when no weather data is available:
    \begin{align}
        \mu\left(w_i,T_i;hrz0\right) \rightarrow \tilde{\mu}
    \end{align}
    in \eqref{eq:stochastic_cumulative_soiled_area} (i.e. we ignore the mean prediction of the physical model). The measured reflectance loss in \eqref{eq:reflectance_difference} will therefore be distributed as
    \begin{align}
        \Delta r_{\ell,k_i} & \sim \mathcal{N}\left(\tilde{\mu}_{\ell,k_i,k_{i-1}},\sigma_{\ell,k_i,k_{i-1}}^2\right) \label{eq:difference_distribution_constant_mean}
    \end{align}
    with
        \begin{align}
            \tilde{\mu}_{\ell,k_i,k_{i-1}}    & =  - \tilde{\mu} \cdot b(\phi_{k_i})\sum_{j=k_{i-1}}^{k_i-1} \alpha_j\,\cos(\theta_j) \label{eq:loss_mean_constant}
        \end{align}
    and the same variance equation \eqref{eq:loss_variance} still holds. In this case, the deposition and Mie loss components of the physical model are absorbed into a single average and only the incidence angle, tilt angle, and the dust scaling factor are needed as inputs.  This model was also estimated by maximizing \eqref{eq:likelihood} with $\tilde{\mu}_{\ell,k_i,k_{i-1}}$ in place of $\mu_{\ell,k_i,k_{i-1}}$. For the estimation of this simplified model, $\log\left( \tilde{\mu} \right)$ replaces $\log\left[ \log\left( hrz0 \right) \right]$ as a decision variable.
\subsection{Statistical distribution of daily losses for the constant-mean model}
    \label{sec:daily_loss_statistics}
    The constant-mean model permits a simple procedure for computing the daily reflectance loss distribution, which is a key practical parameter for CSP. The procedure described here can be extend in the case of the semi-physical model though straightforward modifications, albeit with a more complex simulation procedure.   
    
    For the constant-mean model, if the parameter uncertainty is neglected, the statistical distribution of losses can be computed via \eqref{eq:difference_distribution_constant_mean}, but with $\sigma_{r,k_{i}} = \sigma_{r,k_{i-1}} = 0$ in order to predict the ``true'' reflectance instead of the noise-polluted measurements. Note that this statistical distribution is dependent on the total dust loading through the sums:
    \begin{align}
        & \sum_{j=k_{i-1}}^{k_i-1} \alpha_j\,\cos(\theta_j) \label{eq:dust_loading} \\
        & \sum_{j=k_{i-1}}^{k_i-1} \alpha_j^2 \,\cos^2(\theta_j)
    \end{align}
    which provide convenient metrics for identifying the dust loading of a particular time interval (e.g. one day)\footnote{These loadings are not sufficient for the semi-physical model, where deposition has some complex dependencies on the wind speed and ambient temperature as well as the airborne dust. For this model, the mean loss in \eqref{eq:loss_mean} is the only meaningful quantification of the loading.}.  Including the parameter uncertainty is also straightforward if one is willing to resort to Monte Carlo simulation, since $\log(\tilde{\mu})$ and $\log(\sigma_{dep})$ are approximately jointly normal under MLE. A simple Monte Carlo simulation can thus be conducted to obtain the reflectance changes for any day as follows:
    \begin{enumerate}[Step 1.]
        \item \label{list:dust_loading_computation} Compute the dust loading sums $\sum_{j=k_{i-1}}^{k_i-1} \alpha_j\,\cos(\theta_j)$, and \newline $\sum_{j=k_{i-1}}^{k_i-1} \alpha_j^2 \,\cos^2(\theta_j)$ for all days where airborne dust is available (but not necessarily reflectance measurements). Here, $k_{i-1}$ is the index of the first interval for the day and $k_i$ is the index of the last measurement interval.
        \item \label{list:monte_carlo_first_step} Sample the daily dust loading from those computed in step \ref{list:dust_loading_computation}
        \item \label{list:parameter_sampling} Sample $\log(\tilde{\mu})$ and $\log(\sigma_{dep})$ from a joint normal
        \item Compute the mean and variance via \eqref{eq:loss_variance},\eqref{eq:loss_mean_constant} with $\sigma_{r,k_{i}} = \sigma_{r,k_{i-1}} = 0$
        \item \label{list:monte_carlo_last_step} Sample reflectance changes \eqref{eq:difference_distribution_constant_mean}
        \item Repeat steps \ref{list:monte_carlo_first_step} to \ref{list:monte_carlo_last_step} until the desired number of samples have been obtained.
    \end{enumerate}
    The statistics of the obtained samples can be used to estimate the characteristics of the daily losses. 
    

%% file: sections/results.tex
The accuracy of the model described in Section~\ref{sec:modeling} was assessed against experimental data collected at three locations in Australia. The parameters ($hrz0$, $\sigma_{dep}$, $\tilde{\mu}$) were fitted using the data and the measured mirror reflectance was compared with the simulated values, including analysis of their uncertainties. Ten approximately week-long experimental campaigns conducted at three different sites were exploited to fit and test the model predictions. Four of the campaigns were conducted at the first site --- the roof of a building at the Queensland University of Technology (QUT), Gardens Point campus, located in Brisbane. Three campaigns were conducted at the second site, which was near the remote outback mining town of Mount Isa, Queensland, and another three campaigns were conducted at the third site near a factory in Wodonga, Victoria. A diverse range of conditions and weather characteristics were provided by the three environments for testing the predictive performance of the models. All the reflectance measurements collected at the three locations throughout the experimental campaigns have been included in the presented results. The only exception is represented by those measurements performed after rain events that significantly clean the mirrors or when only a very limited time frame (e.g. smaller than 48 hours) was available between rain-induced cleaning, due to the fact that such small losses were well within the uncertainty of the average reflectance measurements.

The experimental setups for the three sites and the data collection procedure are described in the remainder of the section. Subsequently, the fitting of the models is discussed and their predictive performance on novel data sets is examined.

\subsection{Experiment \& model setup}
The three experimental setups were deployed in different years in the two locations, and are clearly exposed to different conditions. However, their conceptual design is similar and they both share the same main components. The two main equipment that are deployed to perform the experiments are 1) the dust sampler and 2) the mirror rig. Two different types of dust sampler were used for the experimental campaigns. The dust sampler used in the experiments conducted at QUT and in Mount Isa is a Protinus 1000 purchased from Ecotech that measures Total Suspend Particles (TSP), and is equipped with additional weather sensors to measure wind speed and direction, relative humidity, and air temperature among others. The dust sampler used in the experiments conducted in Wodonga is a QAMS Dust Master Pro purchased from Thomson Environmental Systems that simultaneously measures five different Particulate Matter (PM) fractions, and is equipped with additional weather sensors to measure wind speed and direction, relative humidity, air temperature, and rain intensity among others. Both dust samplers uses a light scatter method to assess the mass of airborne dust, that may be integrated with a gravimetric analysis on collected samples. The mirror rigs are composed of a few mirror samples tilted at different angles to simulate the changing position of heliostats tracking the sun in an actual solar field. Eventually, a device is required to measure the reflectance of the mirrors, both in clean and soiled state. In all experiments, a multi wavelengths and acceptance angle configurable 15R-RGB reflectometer purchased from Devices \& Services Company (D\&S) was used to collect measurements. The measurements repeatability reported in the technical sheet of the instrument is ±0.2\%. The incidence angle is fixed by the device at $\theta$ = 15$^{\circ}$. Detailed information on the devices can be found in \cite{picotti_development_2018,picotti_evaluation_2021}.

\begin{figure*}[htp]
    \centering    
    \includegraphics[width=0.6\textwidth,keepaspectratio]{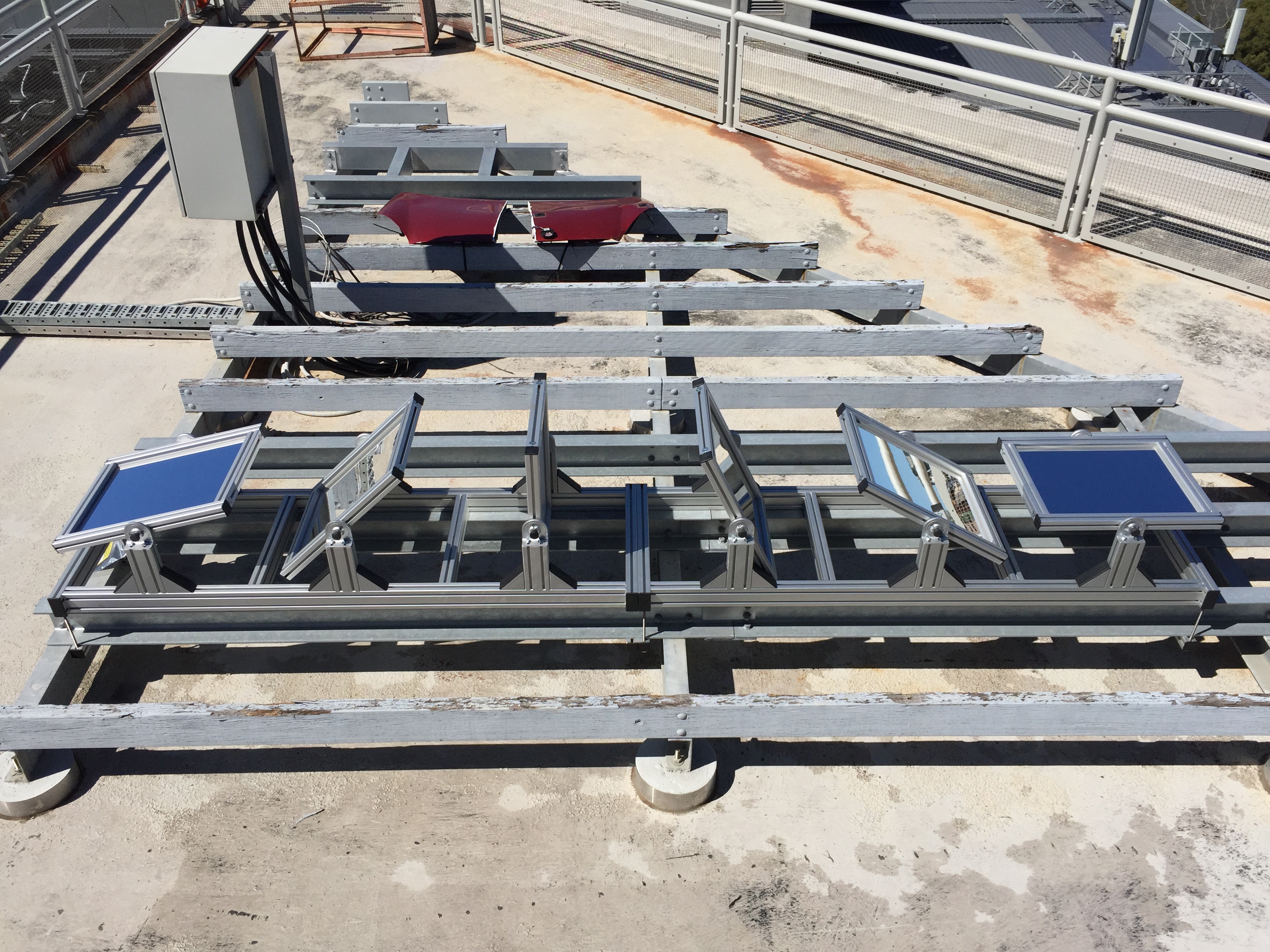}
    \caption{Mirrors Setup at QUT}
    \label{fig:mirrorsqut}
\end{figure*}

The experimental setup at QUT is located on an intermediate level on the roof of a twelve-story building in the CBD of Brisbane, and it is comprised of the dust and weather sampler and a 6 mirrors test-rig oriented at 0$^{\circ}$, 15$^{\circ}$, 30$^{\circ}$, 45$^{\circ}$, 60$^{\circ}$, and 90$^{\circ}$, represented in Fig.~\ref{fig:mirrorsqut}. The settings of the D\&S reflectometer were set on green wavelength ($\lambda$ = 560nm) and an (half) acceptance angle $\theta_a$ = 12.5mrad. Four separate measurement campaigns are available and the details of these campaigns can be found in \cite{picotti_development_2018}.

\begin{figure*}[htp]
    \begin{subfigure}{0.49\textwidth}
        \centering
        \includegraphics[width=0.9\textwidth]{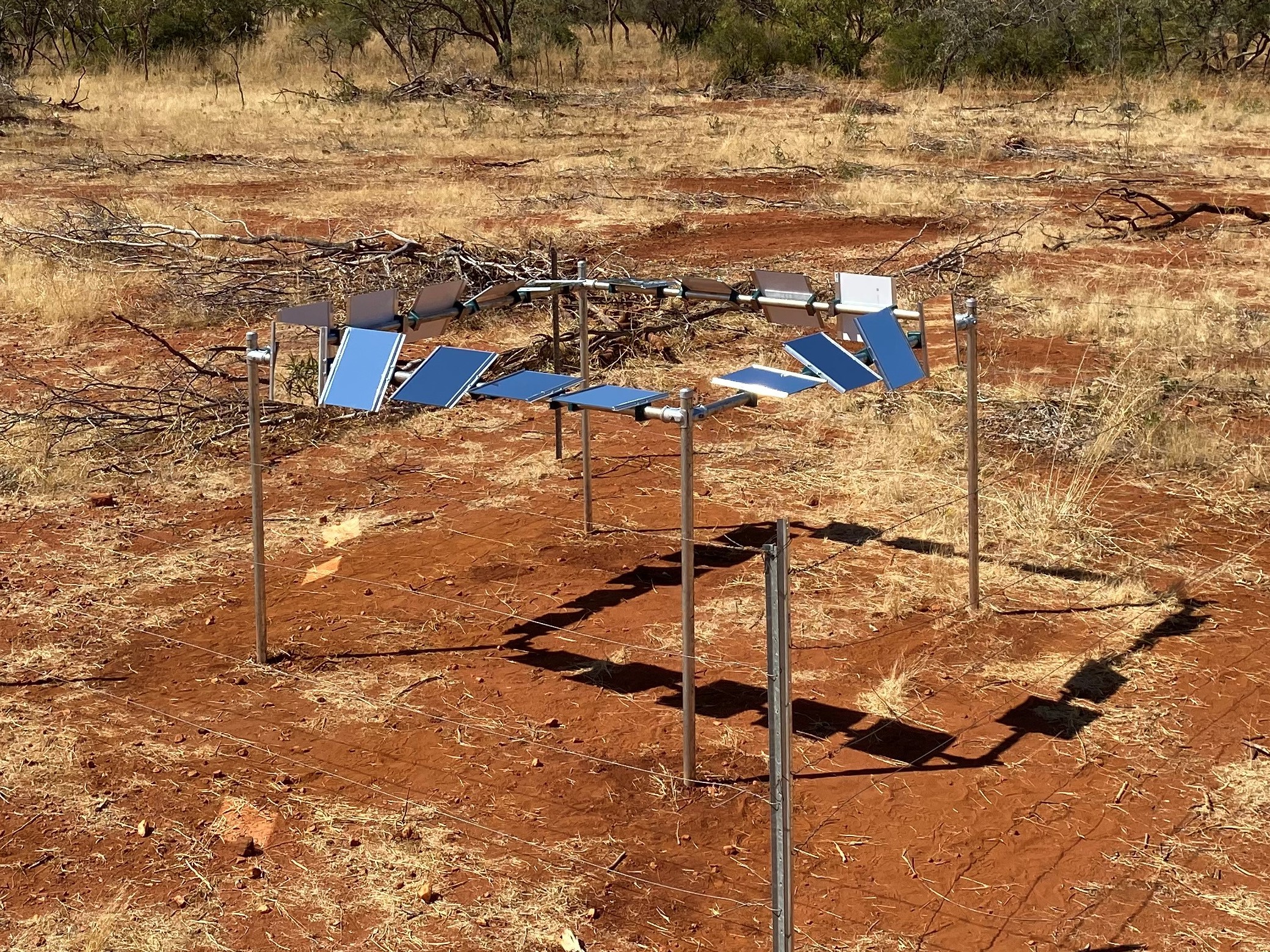}
        \caption{Mirrors test-rig}
        \label{fig:testrig-isa}
    \end{subfigure}
    \hfill
    \begin{subfigure}{0.49\textwidth}
        \centering        
        \includegraphics[angle=0,height=0.85\textwidth]{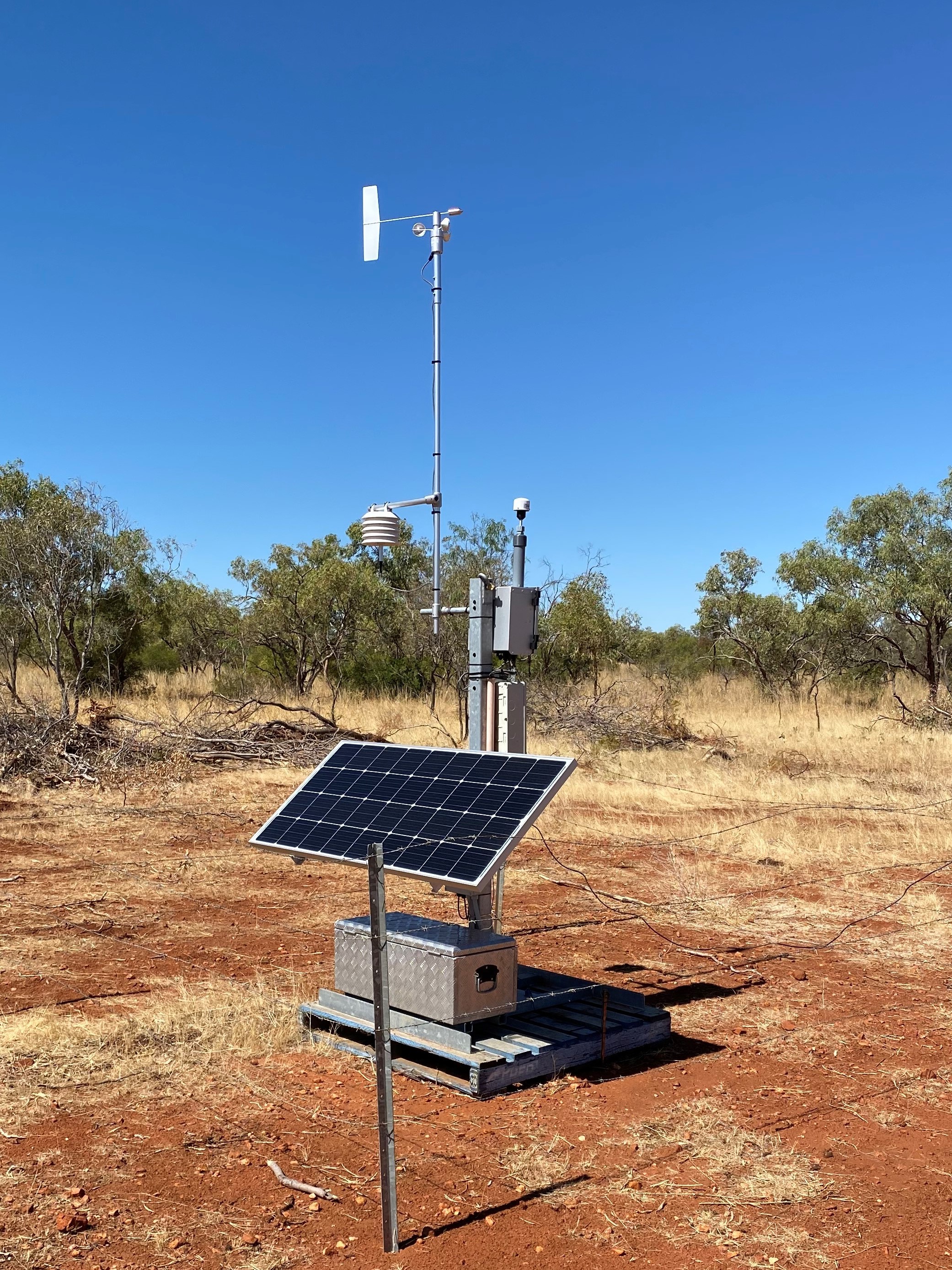}
        \caption{Dust sampler equipped with PV panel}
        \label{fig:dustsampler-isa}
    \end{subfigure}
    \caption{Experimental Setup in the Australian Outback - Details}
    \label{fig:exp-setup-isa-details}
\end{figure*}

The experimental setup in the Australian outback is located a few kilometers away from Mount Isa, and it is surrounded by typical low bush vegetation and red soil. The dust sampler and mirror rig were approximately 20~m apart and fences were constructed for protection from animals (e.g. kangaroos, emus, cows) that inhabit the area. The test-rig was made of a total of 18 mirrors --- four groups of four mirrors oriented facing the major cardinal directions with tilts of 5$^{\circ}$, 30$^{\circ}$, 60$^{\circ}$, and 85$^{\circ}$. Two additional mirrors are added for control purposes: a horizontal (0$^{\circ}$) one in the North arm of the test-rig and a vertical (90$^{\circ}$) one facing East. The D\&S reflectometer was set on red wavelength ($\lambda$ = 650~nm) and an (half) acceptance angle $\theta_a$ = 12.5~mrad. Details of the dust sampler and the mirror test-rig are visible in Figs.~\ref{fig:testrig-isa} and~\ref{fig:dustsampler-isa}. Three separate experimental campaigns were conducted where the reflectance was measured twice daily using an average of nine different positions on each sample mirror.  The measurement frequency was chosen to balance among required time, dataset size, and detectable loss between measurements.

\FloatBarrier

\begin{figure*}[htbp]
    \begin{subfigure}{0.48\textwidth}
        \centering
        \includegraphics[width=0.9\textwidth]{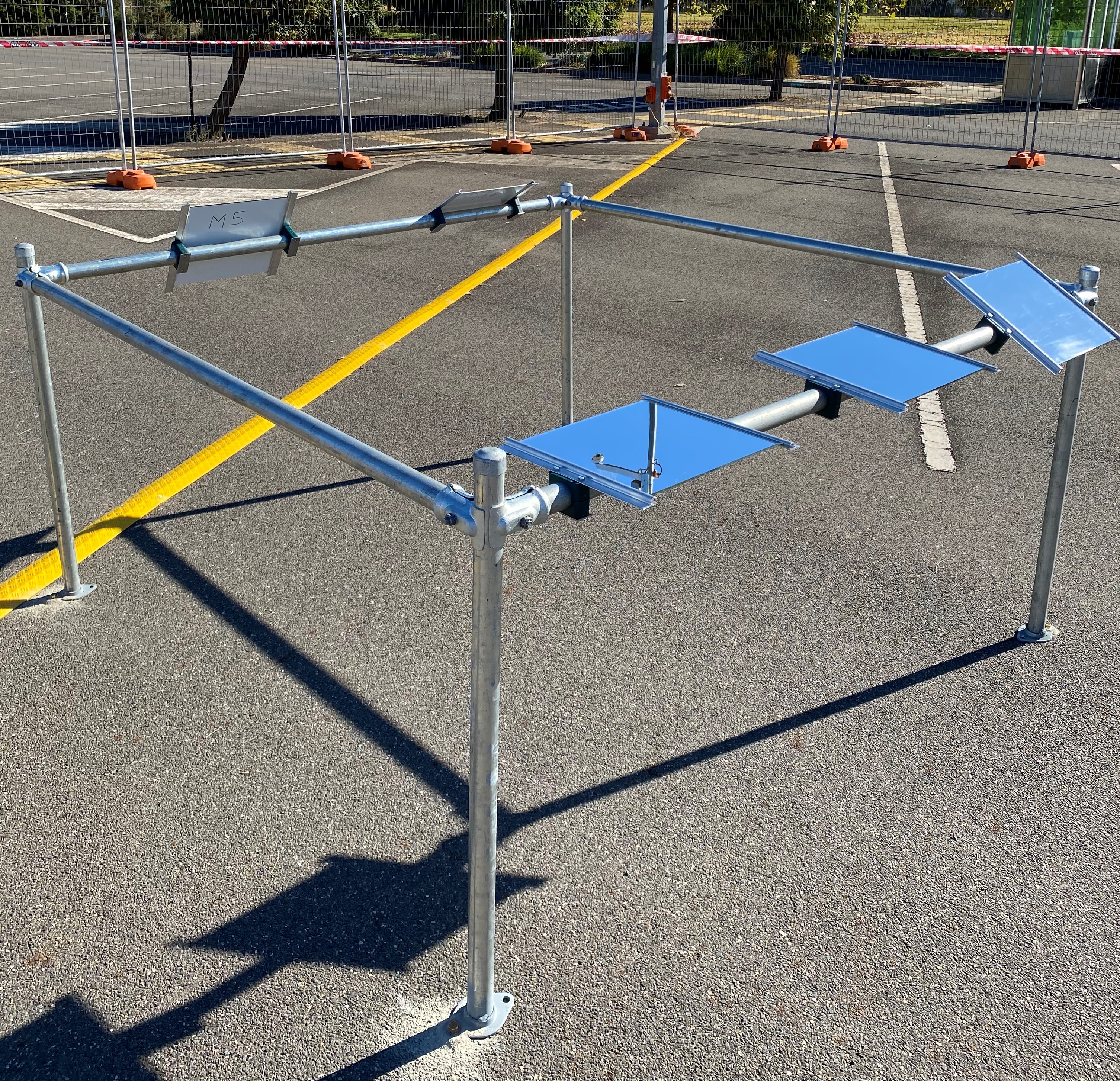}
        \caption{Mirrors test-rig}
        \label{fig:testrig-wodonga}
    \end{subfigure}
    \hfill
    \begin{subfigure}{0.4\textwidth}
        \centering        
        \includegraphics[angle=0,height=0.8\textwidth,angle=-90]{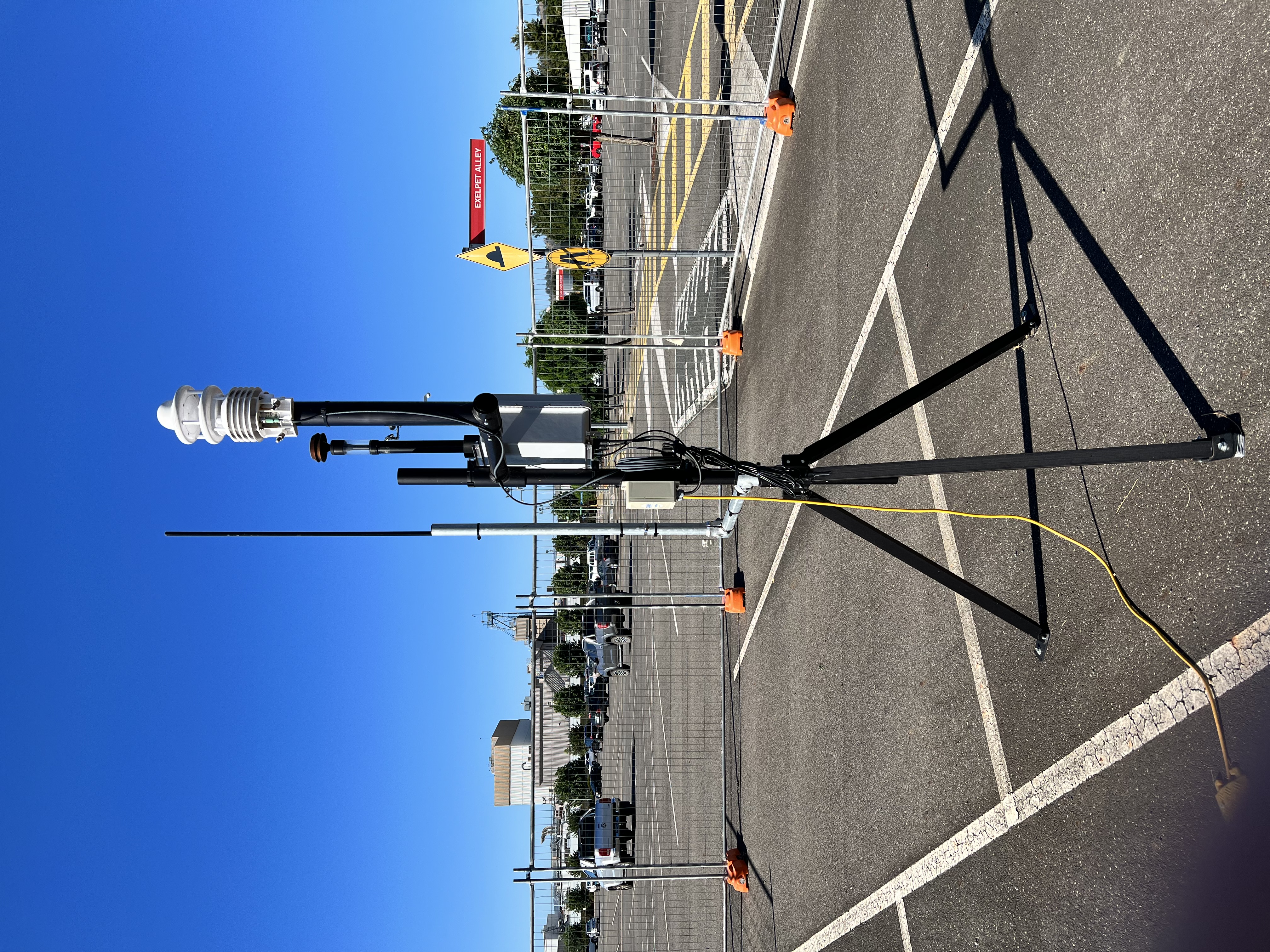}
        \caption{Dust sampler}
        \label{fig:dustsampler-wodonga}
    \end{subfigure}
    \caption{Experimental Setup in Rural Australia - Details}
    \label{fig:exp-setup-wodonga-details}
\end{figure*}

The experimental setup in Wodonga, Victoria, is located in the parking lot of a factory in the outskirts of the town. It is comprised of a mirror test-rig, a dust sampler, and a weather station set up. The mirror rig is made of 5 mirrors deployed at different tilt angles: M1 –- 0$^{\circ}$, M2 -– 5$^{\circ}$, M3 -– 30$^{\circ}$, M4 –- 30$^{\circ}$, and M5 -– 60$^{\circ}$, facing East (M1 to M3) and West (M4 and M5). The weather station measured temperature, relative humidity, ambient pressure, wind speed, wind direction, and precipitation intensity, among other parameters. The dust sampler measured five fractions of PM (Particulate Matter): PM1, PM2.5, PM4, PM10, and PM20.  The settings of the D\&S reflectometer were set on red wavelength ($\lambda$ = 650~nm) and a (half) acceptance angle of $\theta_a$ = 12.5~mrad. Details of the dust sampler and the mirror test-rig are visible in Figs.~\ref{fig:testrig-wodonga} and~\ref{fig:dustsampler-wodonga}. Three separate experimental campaigns were conducted where the reflectance was measured twice a day sampling nine different positions on each sample mirror and taking the average of those measurements.
\FloatBarrier
For the semi-physical model, a number of parameters regarding the size and composition of airborne dust must be specified. The airborne dust distributions were assumed based on those from \cite{picotti_development_2018}: the Urban prototype was used for the QUT experiments and the Rural prototype was used for both Mount Isa and Wodonga. In all cases, the dust was assumed to be composed of homogeneous quartz spheres. The (half) acceptance angle of the reflectometer was 12.5 mrad and the incidence angle was 15$^\circ$. Other parameter values were set to those in Appendix B of \cite{picotti_development_2018}.

Both the semi-physical and constant-mean models were implemented in Python and are now available in the HelioSoil GitHub repository\footnote{https://github.com/cholette/HelioSoil}. The implementations rely mostly on NumPy\cite{harris_array_2020} and SciPy\cite{virtanen_scipy_2020}, but the maximization of the log-likelihood is carried out via the minimization of the negative log-likelihood using {\tt scipy.optimize.minimize}. Mie extinction coefficients are computed using {\tt miepython} \cite{prahl_miepython_2023}. The data for all experimental campaigns, model inputs, and model parameter values can also be found in in the {\tt mirror\_soiling\_data} GitHub repository\footnote{https://github.com/cholette/mirror\_soiling\_data}.  

\subsection{Fitting and prediction results}
\begin{figure*}[htp]
    \centering    
    \includegraphics[width=0.95\textwidth,keepaspectratio]{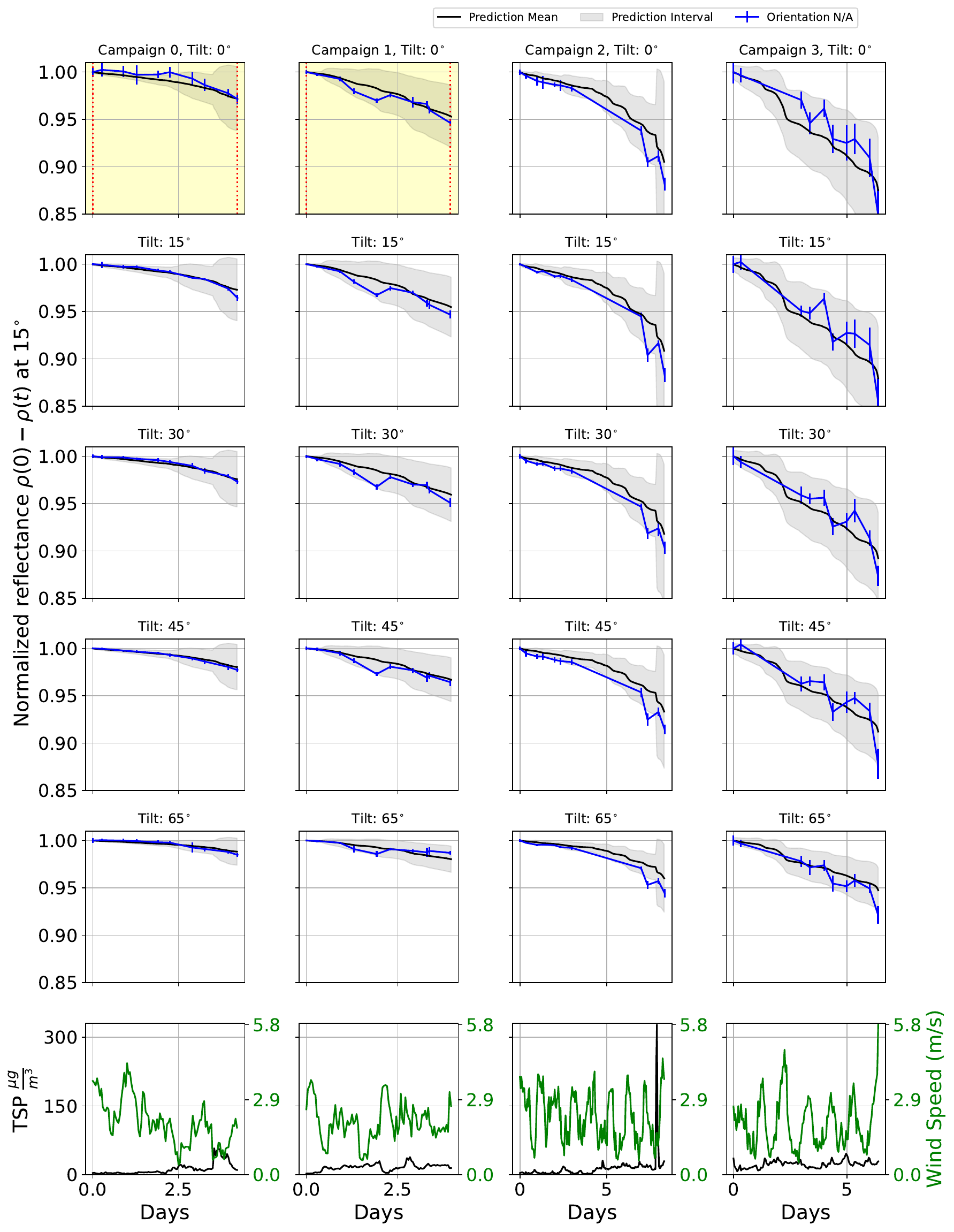}
    \caption{Results for the \textbf{semi-physical model} for the QUT experiments. The reflectance measurements and predictions have been shifted to start at 1.0 (and hence labelled ``normalized reflectance''). The yellow shading denotes data used for fitting.}
    \label{fig:qut_results_physical}
\end{figure*}
\begin{figure*}[htp]
    \centering    
    \includegraphics[width=0.95\textwidth,keepaspectratio]{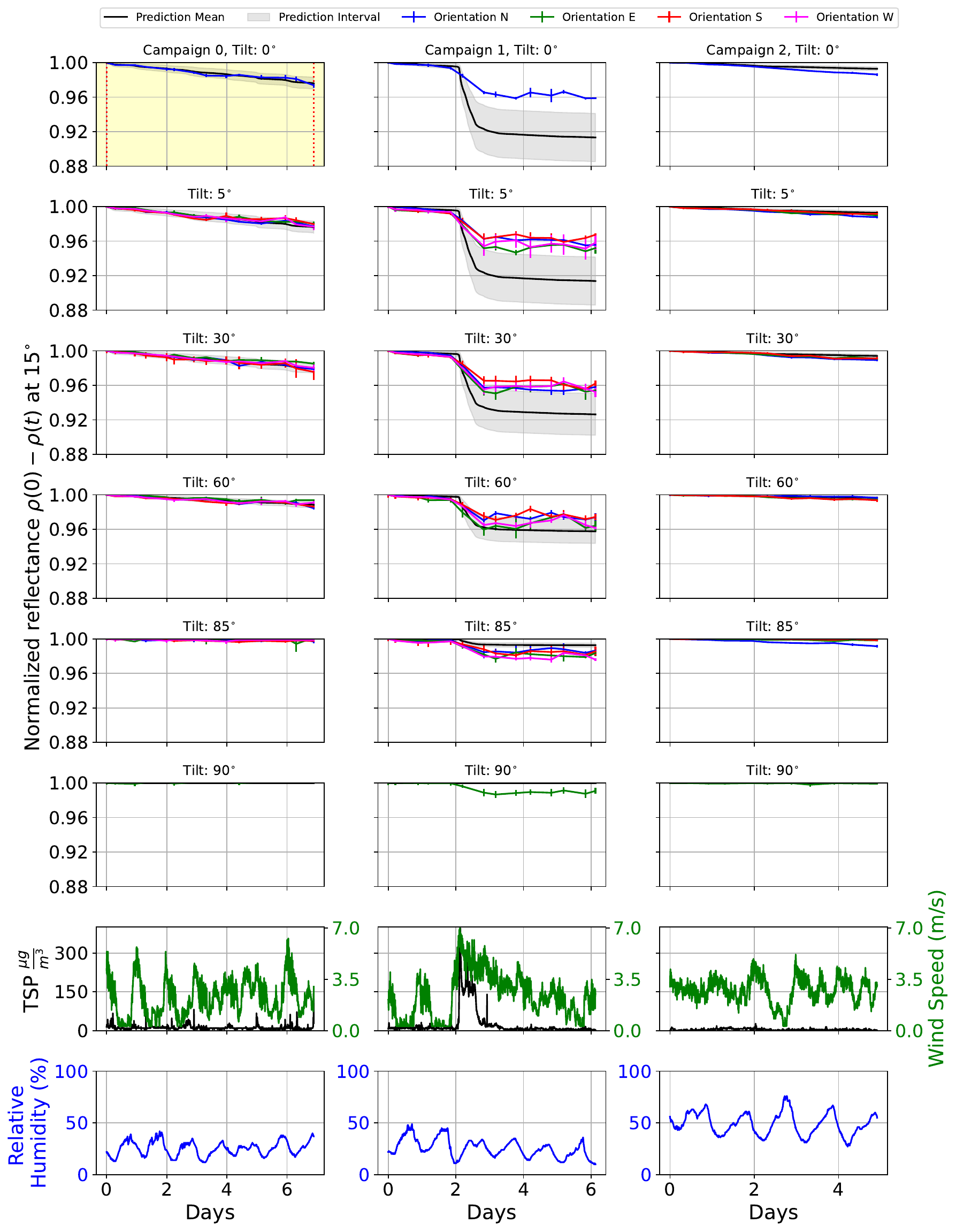}
    \caption{Results for the \textbf{semi-physical model} for the Mount Isa experiments. The reflectance measurements and predictions have been shifted to start at 1.0 (and hence labelled ``normalized reflectance''). The yellow shading denotes data used for fitting.}
    \label{fig:mount_isa_results_physical}
\end{figure*}
\begin{figure*}[htp]
    \centering    
    \includegraphics[width=0.95\textwidth,keepaspectratio]{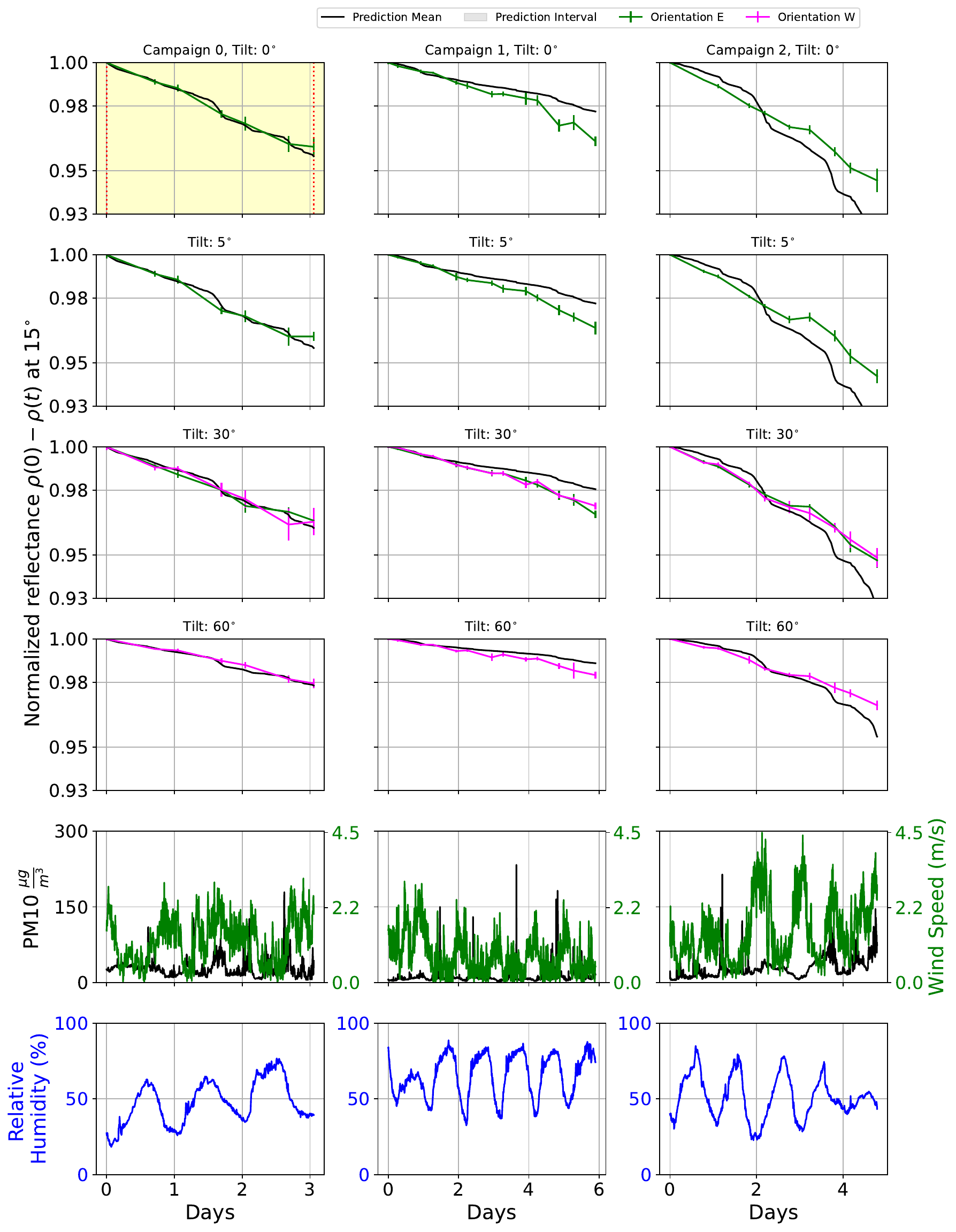}
    \caption{Results for the \textbf{semi-physical model} for the Wodonga experiments. The reflectance measurements and predictions have been shifted to start at 1.0 (and hence labelled ``normalized reflectance''). The yellow shading denotes data used for fitting.}
    \label{fig:wodonga_results_physical}
\end{figure*}

For each site, the parameters of the models were estimated using data from the horizontal mirrors. For the QUT experiments, the first 2 (out of 4) campaigns were used for fitting, while the first campaign (out of three) was used for Mount Isa and Wodonga. For Wodonga, two additional fittings were performed using two and three campaigns to assess the impact of including more data. A detailed analysis of the prediction quality will be presented in this section and will focus on the ability of the model to predict mean losses and estimate prediction confidence intervals. In addition, a more traditional analysis using root-mean-square error and visual quality checks is presented in~\ref{appendix_A}.
\begin{table}[htbp]
    \centering
    \caption{Maximum likelihood estimates and 95\% confidence interval for the \textbf{semi-physical model}.}
    \label{tab:semi-physical_model_fitting}
    \begin{tabular}{c|ccc}
            \textbf{Location}   & $\Delta t$   & $\widehat{\mathbf{hrz0}}$   & $\hat{\boldsymbol{\sigma}}_{dep} \times 10^{4}$ \\
        \hline
        QUT                     & 1 hr         & $1.88 \; \left[1.51,\, 2.63\right]$   & $2.90 \; \left[1.74,\, 4.81\right]$   \\
        Mount Isa               & 5 min        & $7.11 \; \left[3.43,\, 22.6\right]$   &  $1.40 \; \left[0.400,\, 4.90\right]$  \\
        Wodonga (one exp.)      & 5 min        & $2.71 \; \left[2.39,\, 3.12\right]$   & $\sim 0 \; \left[0.00, \,\infty\right]$  \\
        Wodonga (two exp.)      & 5 min        & $3.31 \; \left[2.18,\, 6.26\right]$   & $1.26 \; \left[0.749, \, 2.13\right]$  \\
        Wodonga (three exp.)    & 5 min        & $5.61 \; \left[2.97,\, 15.3\right]$   & $1.49 \; \left[1.05, \, 2.11\right]$  \\
    \end{tabular}
\end{table}
The results of the semi-physical model fitting can be seen in Table \ref{tab:semi-physical_model_fitting} and the model performance can be seen in Figs.~\ref{fig:qut_results_physical}, \ref{fig:mount_isa_results_physical}, and \ref{fig:wodonga_results_physical} for the QUT, Mount Isa, and Wondonga experiments, respectively. In these plots, each column of subplots shows a different experimental campaign, while each row represents a mirror with a different tilt. The black lines denote the predictions of the model from the first measurement onward, while the grey shaded areas indicate the two-standard-deviation prediction interval. The remaining lines denote the mean of nine reflectance measurements with the error bars denoting the sample standard deviation of this mean. The yellow shading denotes time periods used for fitting the model parameters.
The QUT experiments can be found in Fig.~\ref{fig:qut_results_physical}. A few observations may be made regarding the training experiments (the first two columns of subplots): 1) the model describes the effect of increasing tilt without the need for training on additional angles; 2) the width of the prediction intervals (grey shaded areas) increases with increasing airborne dust concentration. This is most evident when the TSP increased toward the end of Experiment 0, where a corresponding increase in the growth rate of the prediction interval is observed. The validation experiments (the second two columns of subplots in Fig.~\ref{fig:qut_results_physical}) demonstrate good agreement between the model and experimental measurements.

The semi-physical model results for the Mount Isa experiments can be seen in Fig.~\ref{fig:mount_isa_results_physical}. In this experiment, there were a number of mirrors with different orientations to evaluate the influence of wind direction, hence the additional experimental curves present (particularly at 5-80 degrees of tilt). Note that for the visual comparison of the reflectance degradation trends, the initial reflectances (which differed by about 1 percentage point) were all shifted upward so that they have a common starting point (1.0). Interestingly, all orientations exhibit similar losses for the same tilts (rows in Fig.~\ref{fig:mount_isa_results_physical}), with the West and East orientations perhaps showing higher losses at the end of the test. However, this difference in losses is small and well within the measurement mean error bars.

In the fitting campaign, the effect of the tilt angle is modelled well once again. Moreover, third campaign shows excellent agreement with the model predictions using the parameter fit from the first campaign, however the losses are quite low. The results for the second campaign are mixed. A high dust event was observed around Day 2 (TSP near 300$\frac{\mu g}{m^3}$) and the losses are clearly over-predicted by the model, even if the model does communicate a lack of confidence in the predictions through a significant expansion of the prediction interval. Both before and after this event, the loss rate is still predicted quite well by the model, though this loss rate is quite low again. Interestingly, the losses observed on the vertical (90$^\circ$) and near-vertical (85$^\circ$) mirrors, which indicate the presence of a horizontal deposition mode that is neglected in both models. This may be the culprit for the trend from overestimation of losses for flat mirrors toward underestimation of losses for the vertical mirror during the high-dust event. Finally, the losses don't seem to show significant orientation dependence.

\begin{figure*}[htp]
    \centering    
    \includegraphics[width=0.95\textwidth,keepaspectratio]{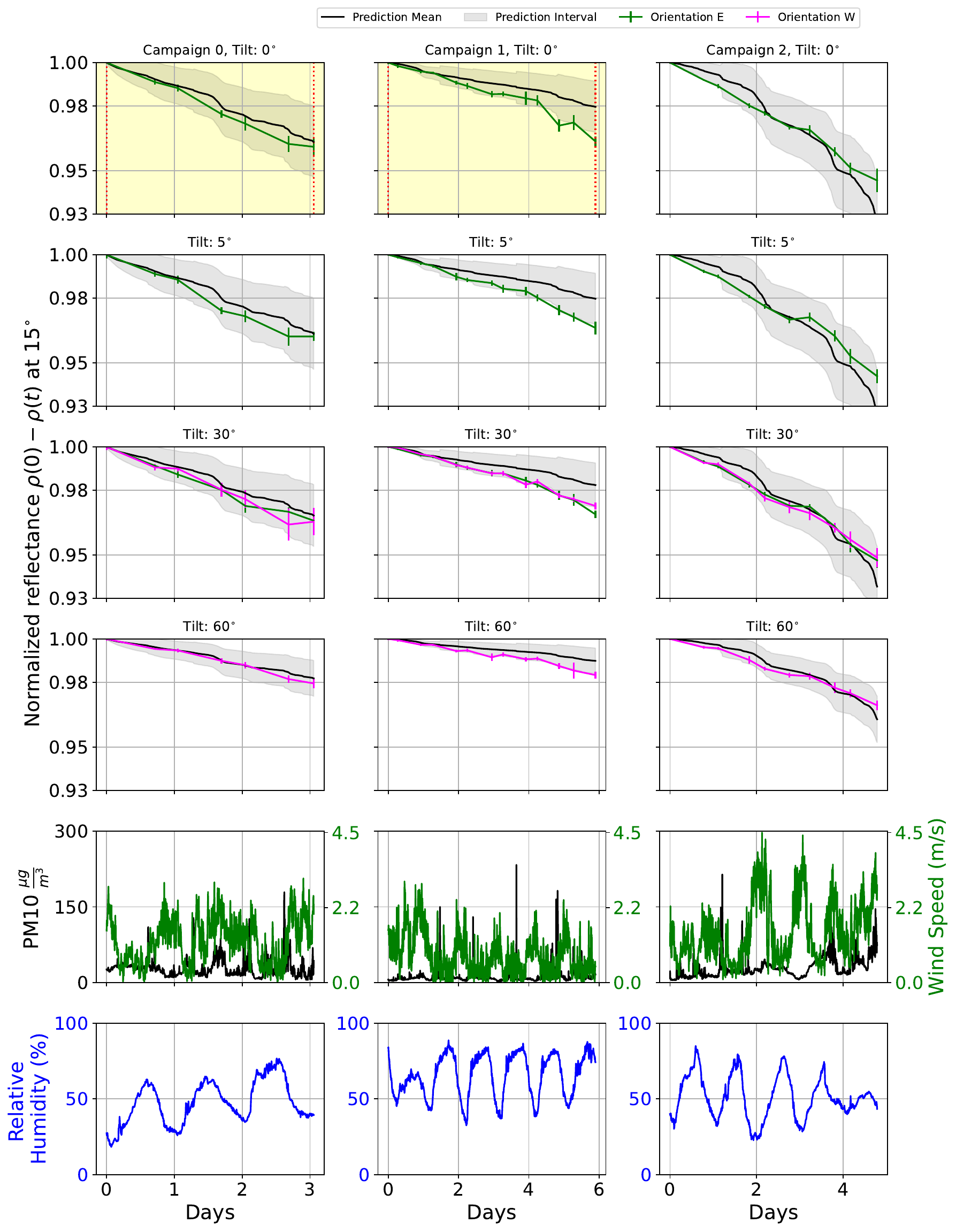}
    \caption{Results for the \textbf{semi-physical model} for Wodonga, but now trained using the first two experiments. The reflectance measurements and predictions have been shifted to start at 1.0 (and hence labelled ``normalized reflectance''). The yellow shading denotes data used for fitting.}
    \label{fig:wodonga_results_physical_2}
\end{figure*}

The results for the Wodonga experiments can be seen in Figs.~\ref{fig:wodonga_results_physical} and \ref{fig:wodonga_results_physical_2}. Initial reflectances were again all shifted up so that they have a common starting point (1.0) for visual comparison. The semi-physical model appears to capture the tilt angle effects in the first campaign, but testing on the other two campaigns yields poor performance. Moreover, the estimated $\sigma_{dep}$ is very small, which is the reason that no confidence interval is visible in the plots (it is too small to be seen). However, the confidence interval for this parameter is trivially large due to a poorly conditioned Hessian at the maximum, clearly indicating that one should be suspicious of this estimate. Adding the second experiment to the training set improves the confidence interval on $\sigma_{dep}$, as can be seen in Table \ref{tab:semi-physical_model_fitting}. However, there are two indications here that the model is missing something: 1) the $hrz0$ value for the two-experiment fit is outside the confidence interval of the one-experiment fit and 2) the fitting and prediction results seen in Fig. \ref{fig:wodonga_results_physical_2} show that the fit is still poor on the second experiment, despite the fact that it was used in the parameter estimation. When all three experiments are used (Table \ref{tab:semi-physical_model_fitting}) --- $hrz0$ increases, albeit inside the two-experiment confidence interval this time, but the confidence interval expands to the right. Taken together, the three Wodonga fittings suggest that there are some phenomena that are driving the differences in soiling rates between the experiments that are not captured by the semi-physical model. 

The fitting results for the constant-mean models can be seen in Table~\ref{tab:constant_mean_fitting}.
\begin{table}[htbp]
    \centering
    \caption{Maximum likelihood estimates and 95\% confidence interval for the \textbf{constant-mean model}.}
    \label{tab:constant_mean_fitting}
    \begin{tabular}{c|ccc}
        \textbf{Location}       & $\Delta t$   & $\hat{\tilde{\boldsymbol{\mu}}} \times 10^{4}$    & $\hat{\boldsymbol{\sigma}}_{dep}\times 10^{4}$ \\
        \hline
        QUT                     & 1 hr     & $0.957 \; \left[0.560,\, 1.63  \right]$   & $2.68  \; \left[1.57,  \, 4.57 \right]$   \\
        Mount Isa               & 5 min    & $0.250 \; \left[0.148,\, 0.425 \right]$   & $1.80  \; \left[0.760, \, 4.25 \right]$  \\
        Wodonga (one exp.)      & 5 min    & $0.241 \; \left[0.198,\, 0.293 \right]$   & $0.402 \; \left[0.126, \, 1.29 \right]$  \\
        Wodonga (two exp.)      & 5 min    & $0.261 \; \left[0.218,\, 0.313 \right]$   & $0.641 \; \left[0.298, \, 1.38 \right]$  \\
        Wodonga (three exp.)    & 5 min    & $0.250 \; \left[0.212,\, 0.295 \right]$   & $0.886 \; \left[0.574, \, 1.37 \right]$  \\
    \end{tabular}
\end{table}

\begin{figure*}[htp]
    \centering    
    \includegraphics[width=0.95\textwidth,keepaspectratio]{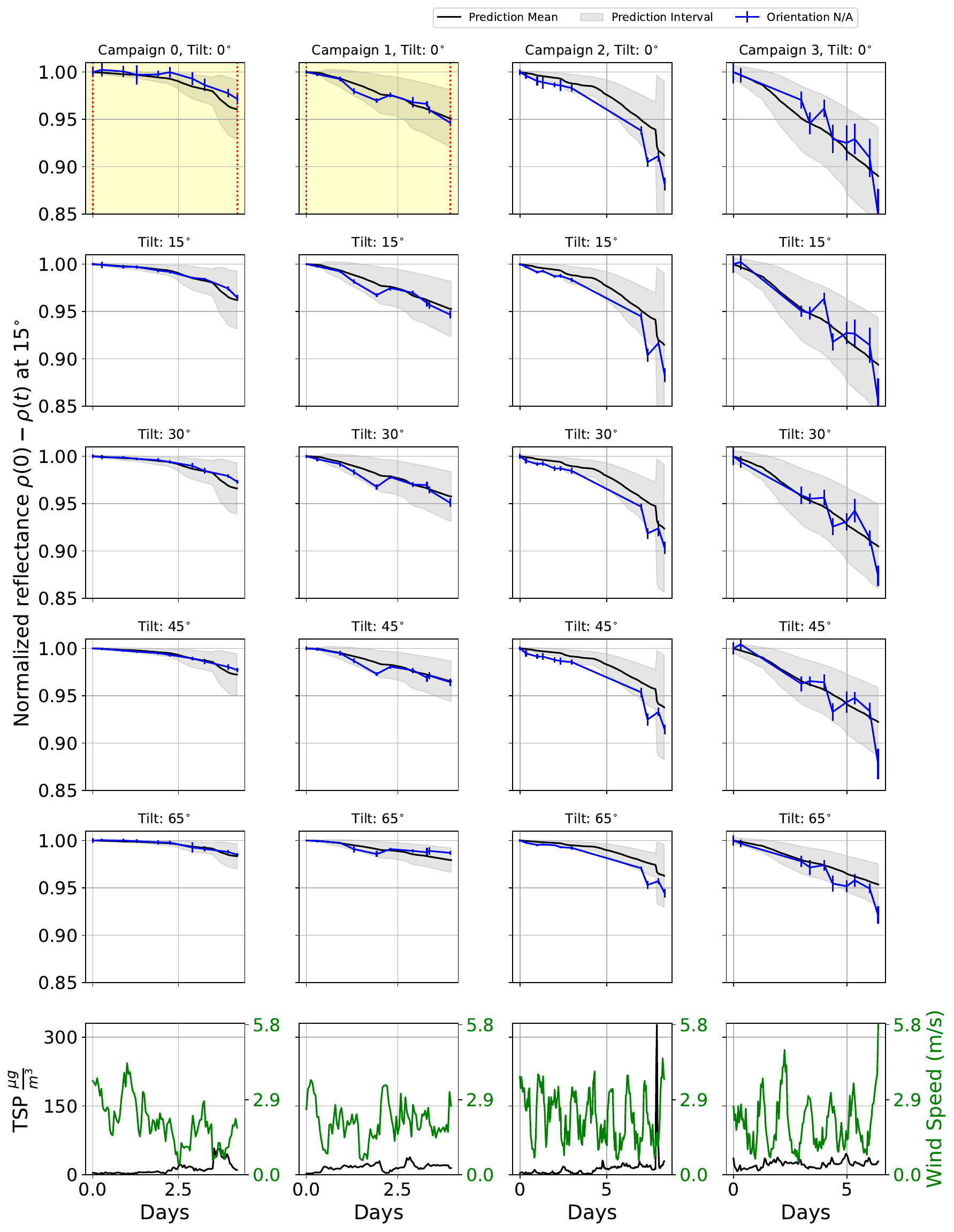}
    \caption{Results for the simplified \textbf{constant-mean} model for the QUT experiments. The reflectance measurements and predictions have been shifted to start at 1.0 (and hence labelled ``normalized reflectance''). The yellow shading denotes data used for fitting.}
    \label{fig:qut_results_simplified}
\end{figure*}

\begin{figure*}[htp]
    \centering    
    \includegraphics[width=0.95\textwidth,keepaspectratio]{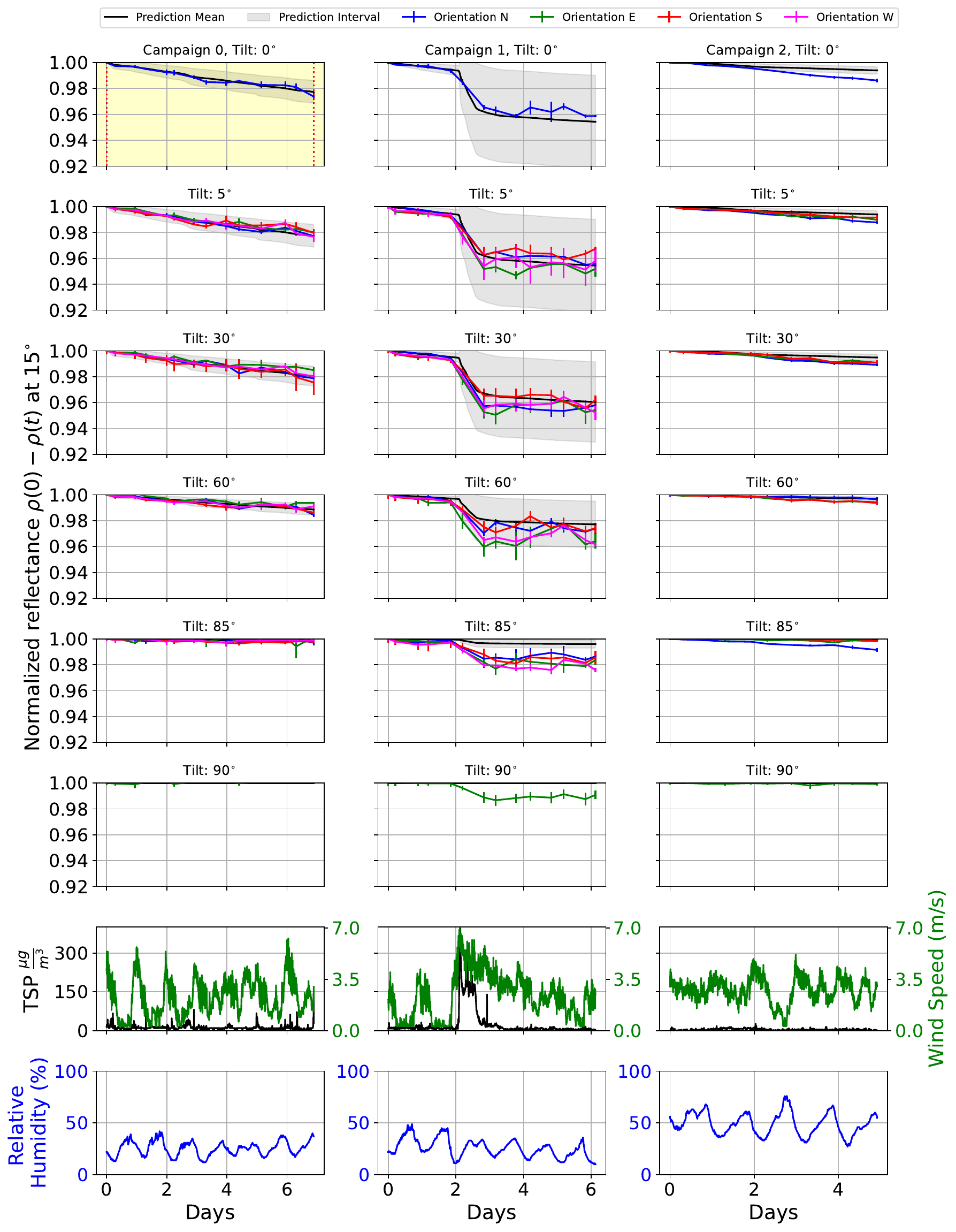}
    \caption{Results for the simplified \textbf{constant-mean} model for the Mount Isa experiments. The reflectance measurements and predictions have been shifted to start at 1.0 (and hence labelled ``normalized reflectance''). The yellow shading denotes data used for fitting.}
    \label{fig:mount_isa_results_simplified}
\end{figure*}

The results of the simplified (constant-mean) model on the QUT experiments are shown in Fig.~\ref{fig:qut_results_simplified} which shows a strikingly similar performance on both the training data and validation experiments to the semi-physical model. The one interesting area of difference is the performance on the final day of the first experiment --- the constant-mean model slightly over-predicts the losses while the semi-physical model does not. This is likely due to the simplified model ignoring the wind speed, which increases during this period. On the other hand, the semi-physical model fits nicely to this behaviour. 
The results for the simplified model on the Mount Isa experiments can be seen in Fig.~\ref{fig:mount_isa_results_simplified}. The performance of this model on the first and third campaign is similar to the semi-physical model. However, the simplified model performs significantly better on the second campaign --- it captures the average behaviour during the high-dust event quite well. There is a hint of some bias in this prediction starting at the 60$^\circ$ tilt, perhaps owing the to the aforementioned apparent ``horizontal'' deposition mode during the high-dust event. 

\begin{figure*}[htp]
    \centering    
    \includegraphics[width=0.95\textwidth,keepaspectratio]{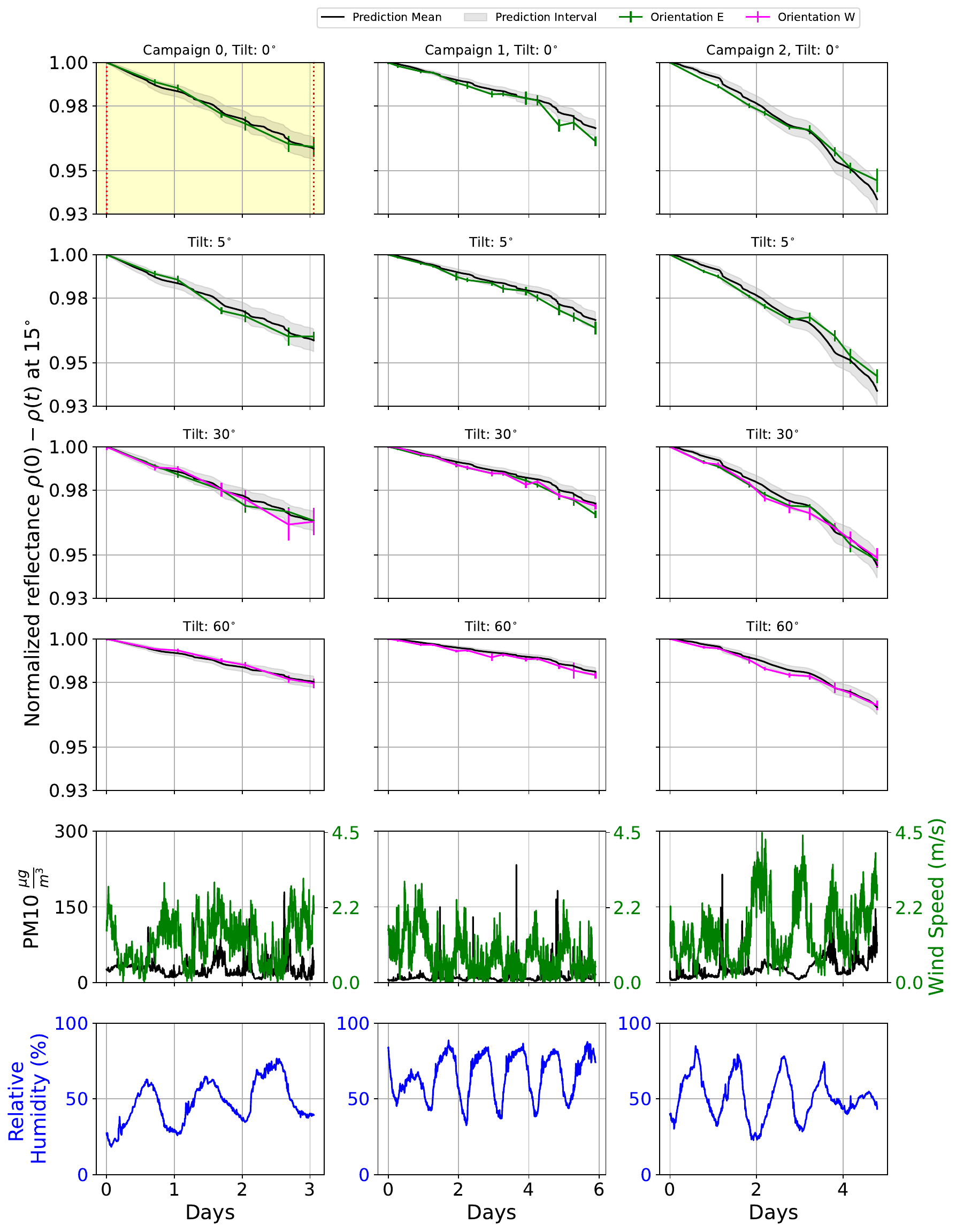}
    \caption{Results for the \textbf{constant-mean} for the Wodonga experiments. The reflectance measurements and predictions have been shifted to start at 1.0 (and hence labelled ``normalized reflectance''). The yellow shading denotes data used for fitting.}
    \label{fig:wodonga_results_constant_mean}
\end{figure*}

The results for the simplified model on the Wodonga experiments can be seen in Fig.~\ref{fig:wodonga_results_constant_mean}. Again, the simplified model shows good prediction of both the tilt angle effects and the losses on the two validation experiments. However, the prediction intervals (the grey area) appear narrow. As can be seen in Table \ref{tab:constant_mean_fitting}, the confidence interval on $\sigma_{dep}$ is quite wide for the one-experiment fit, and adding experiments tends to increase the value (though still within the confidence interval of the one-experiment fitting). 

\subsection{Statistical distributions of losses at the three sites}
\label{sec:distribution_results}
\begin{figure}[htp]
    \centering
    \begin{subfigure}[b]{0.48\textwidth}
         \centering
         \includegraphics[width=\textwidth]{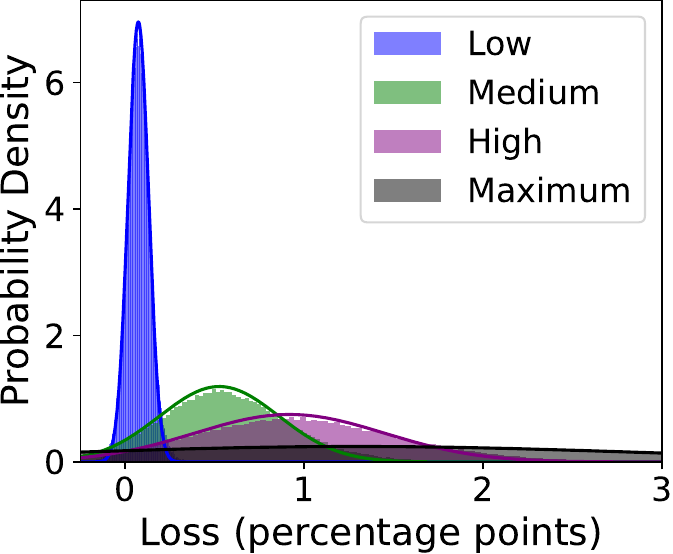}
         \caption{QUT}
         \label{fig:qut_loss_distributions}
     \end{subfigure}
     \hfill \vspace{1ex}
     \begin{subfigure}[b]{0.48\textwidth}
         \centering
         \includegraphics[width=\textwidth]{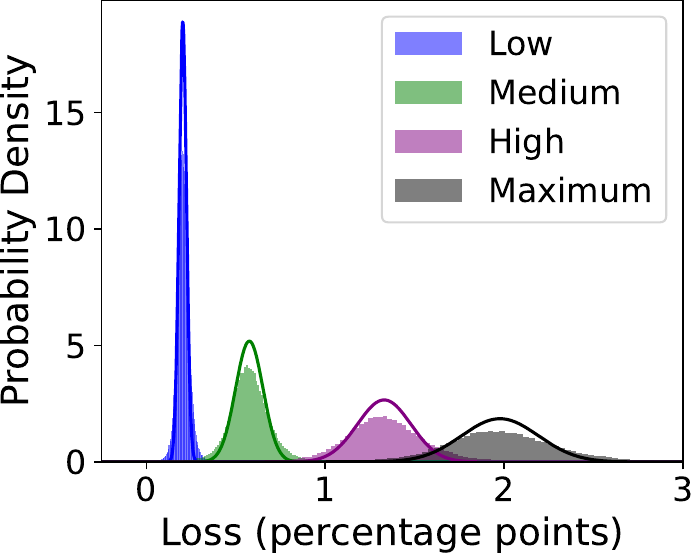}
         \caption{Wodonga}
         \label{fig:wodonga_loss_distributions}
     \end{subfigure}
     \hfill \vspace{1ex}
     \begin{subfigure}[b]{0.48\textwidth}
         \centering
         \includegraphics[width=\textwidth]{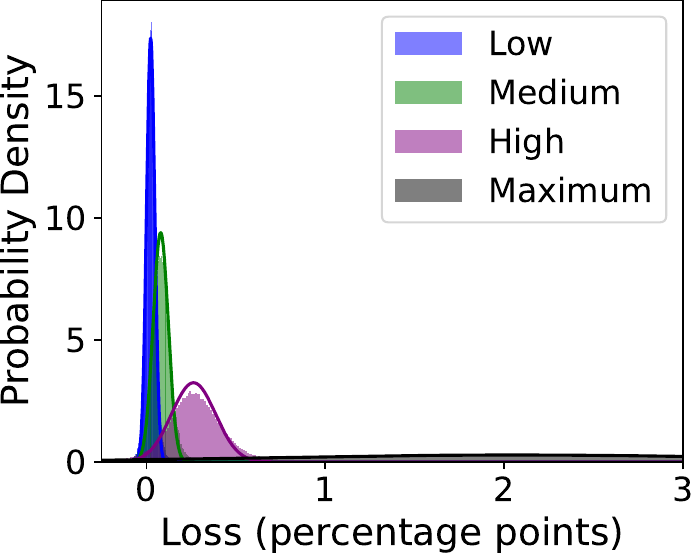}
         \caption{Mount Isa}
         \label{fig:mount_isa_loss_distributions}
     \end{subfigure}
    \hfill
    \begin{subfigure}[b]{0.48\textwidth}
         \centering
         \includegraphics[width=\textwidth]{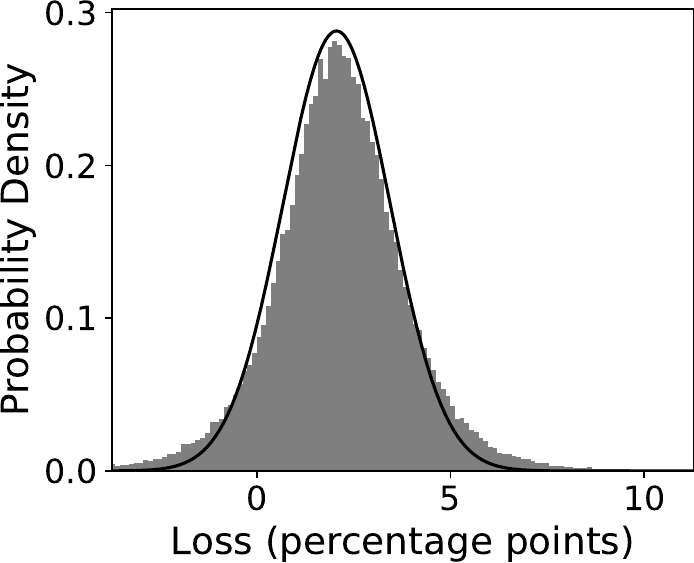}
         \caption{Mount Isa (highest loading only)}
         \label{fig:mount_isa_highest_loss}
     \end{subfigure}
    \caption{Statistical distribution of daily losses for each site under different dust loadings.}
    \label{fig:daily_loss_distributions}
\end{figure}

Using the method described in Section~\ref{sec:daily_loss_statistics}, the statistical distribution of losses for the scenarios was computed for a flat mirror ($\theta_k=0 \; \forall k$) at the three sites. Since the airborne dust data is limited during the experimental campaigns, the sampling of daily loadings in Step \ref{list:dust_loading_computation} is omitted and four dust loading scenarios were considered in its place: low, medium, high, and maximum dust loading scenarios, which are defined as the 5, 50, 95, and 100 \%-ile of the daily sums of \eqref{eq:dust_loading} in the data collected during all of the campaigns. The results are shown in Fig.~\ref{fig:daily_loss_distributions}. The histograms are the result of the Monte Carlo simulations (and therefore include parameter uncertainty), while the solid curves are the analytical distributions using the MLE estimates of the parameters. The losses are summarized in Table \ref{tab:daily_losses} as well. 

\begin{table}[htbp]
    \centering
    \caption{Summary of predicted daily losses (in pp/day) under different dust loadings for each site (mean $\pm$ two standard deviations).}
    \label{tab:daily_losses}
    \begin{tabular}{c|cccc}
        \textbf{Location} & \textbf{Low}    & \textbf{Medium} & \textbf{High}   & \textbf{Maximum} \\
        \hline
        QUT               & $0.07 \pm 0.11$ & $0.53 \pm 0.66$ & $0.92 \pm 1.04$ & $1.29 \pm 3.18$  \\
        Mount Isa         & $0.02 \pm 0.04$ & $0.08 \pm 0.08$ & $0.26 \pm 0.24$ & $2.06 \pm 2.71$  \\
        Wodonga           & $0.20 \pm 0.04$ & $0.58 \pm 0.15$ & $1.33 \pm 0.29$ & $1.98 \pm 0.42$ 
     \end{tabular}
\end{table}
A number of observations can be made from these results. Firstly, typical daily losses during the measurement campaigns (quantified by the medium scenarios) show losses of around 0.5 percentage points per day (pp/day) for both QUT and Wodonga, but is predicted to vary between 0 to 1 pp/day for the same dust loading. For Mount Isa, the typical losses for the observed dust loading is quite low --- around 0.1 pp/day --- and the variance on this distribution is quite low as well. Secondly, the distributions show some probability of a negative loss, which is a result of the normal distribution for the modelling errors. This choice, however, was made to reflect the real possibility of an increase in reflectance due to removal modes that are not included in the physical model modelled (e.g. wind removal).

Thirdly, the variance of the losses increase considerably with dust loading, which is consequence of modelling uncertainty in the deposition velocity. This is most evident in the ``maximum'' scenario in QUT and Mount Isa data --- the very high dust loadings lead to large increases in both the mean and variance of the losses. The Mount Isa highest scenario is plotted separately in Fig.~\ref{fig:mount_isa_highest_loss} to better show this distribution. Note that this dust loading corresponds to the high dust event observed during the second campaign and is responsible for the large increase in the prediction uncertainty observed during this time (see Fig.~\ref{fig:mount_isa_results_simplified}). 

Finally, the relative impact of the three uncertainty sources can be assessed. The differences in the scenario distributions clearly indicate that random fluctuations in the airborne dust has a large impact. Within a scenario, the histograms have slightly higher variance than the curves due to the parameter uncertainty. However, the difference is small, indicating that the within-scenario variance is dominated by uncertainty in the deposition velocity (i.e. $\sigma_{\ell,k_i,k_{i-1}}^2$ ) --- a dominance which will increase if more data is used thereby decreasing the parameter confidence intervals.

%% file: main.bbl
\begin{thebibliography}{10}
\expandafter\ifx\csname url\endcsname\relax
  \def\url#1{\texttt{#1}}\fi
\expandafter\ifx\csname urlprefix\endcsname\relax\def\urlprefix{URL }\fi
\expandafter\ifx\csname href\endcsname\relax
  \def\href#1#2{#2} \def\path#1{#1}\fi

\bibitem{fernandez-garcia_equipment_2017}
A.~{Fern{\'a}ndez-Garc{\'i}a}, F.~Sutter, L.~{Mart{\'i}nez-Arcos}, C.~Sansom,
  F.~Wolfertstetter, C.~Delord, Equipment and methods for measuring reflectance
  of concentrating solar reflector materials, Solar Energy Materials and Solar
  Cells 167 (2017) 28--52.

\bibitem{mehos_concentrating_2020}
M.~Mehos, H.~Price, R.~Cable, D.~Kearney, B.~Kelly, G.~Kolb, F.~Morse,
  Concentrating {{Solar Power Best Practices Study}}, Tech. Rep.
  NREL/TP-5500-75763, {National Renewable Energy Lab. (NREL), Golden, CO
  (United States); Solar Dynamics, LLC, Denver, CO (United States)} (Jun.
  2020).

\bibitem{wales_optimizing_2021}
J.~G. Wales, A.~J. Zolan, A.~M. Newman, M.~J. Wagner, Optimizing vehicle fleet
  and assignment for concentrating solar power plant heliostat washing, IISE
  Transactions 0~(0) (2021) 1--13.

\bibitem{picotti_optimization_2020}
G.~Picotti, L.~Moretti, M.~E. Cholette, M.~Binotti, R.~Simonetti, E.~Martelli,
  T.~A. Steinberg, G.~Manzolini, Optimization of cleaning strategies for
  heliostat fields in solar tower plants, Sol. Energy 204 (2020) 501--514.

\bibitem{terhag_optimization_2019}
F.~Terhag, F.~Wolfertstetter, S.~Wilbert, T.~Hirsch, O.~Schaudt, Optimization
  of cleaning strategies based on {{ANN}} algorithms assessing the benefit of
  soiling rate forecasts, in: {{SOLARPACES}} 2018: {{International Conference}}
  on {{Concentrating Solar Power}} and {{Chemical Energy Systems}},
  {Casablanca, Morocco}, 2019, p. 220005.

\bibitem{truong-ba_sectorial_2020}
H.~{Truong-Ba}, M.~E. Cholette, G.~Picotti, T.~A. Steinberg, G.~Manzolini,
  Sectorial reflectance-based cleaning policy of heliostats for {{Solar Tower}}
  power plants, Renewable Energy 166 (2020) 176--189.

\bibitem{alami_merrouni_csp_2020}
A.~Alami~Merrouni, R.~Concei{\c c}{\~a}o, A.~Mouaky, H.~G. Silva, A.~Ghennioui,
  {{CSP}} performance and yield analysis including soiling measurements for
  {{Morocco}} and {{Portugal}}, Renewable Energy 162 (2020) 1777--1792.

\bibitem{picotti_soiling_2018}
G.~Picotti, P.~Borghesani, M.~Cholette, G.~Manzolini, Soiling of solar
  collectors \textendash{} {{Modelling}} approaches for airborne dust and its
  interactions with surfaces, Renewable and Sustainable Energy Reviews 81
  (2018) 2343--2357.

\bibitem{micheli_economics_2021}
L.~Micheli, E.~F. Fern{\'a}ndez, J.~T. Aguilera, F.~Almonacid, Economics of
  seasonal photovoltaic soiling and cleaning optimization scenarios, Energy 215
  (2021) 119018.

\bibitem{zhu_roadmap_2022}
G.~Zhu, C.~Augustine, R.~Mitchell, M.~Muller, P.~Kurup, A.~Zolan,
  S.~Yellapantula, R.~Brost, K.~Armijo, J.~Sment, R.~Schaller, M.~Gordon,
  M.~Collins, J.~Coventry, J.~Pye, M.~Cholette, G.~Picotti, M.~Arjomandi,
  M.~Emes, D.~Potter, M.~Rae, Roadmap to {{Advance Heliostat Technologies}} for
  {{Concentrating Solar-Thermal Power}}, Tech. Rep. NREL/TP-5700-83041,
  1888029, MainId:83814, {National Renewable Energy Lab. (NREL)} (Sep. 2022).

\bibitem{costa_solar_2018}
S.~C. Costa, A.~S.~A. Diniz, L.~L. Kazmerski, Solar energy dust and soiling
  {{R}}\&{{D}} progress: {{Literature}} review update for 2016, Renewable and
  Sustainable Energy Reviews 82 (2018) 2504--2536.

\bibitem{bonanos_characterization_2020}
A.~M. Bonanos, M.~J. Blanco, K.~Milidonis, Characterization of mirror soiling
  in {{CSP}} applications, in: {{SOLARPACES}} 2019: {{International
  Conference}} on {{Concentrating Solar Power}} and {{Chemical Energy
  Systems}}, {Daegu, South Korea}, 2020, p. 030007.

\bibitem{conceicao_csp_2018}
R.~Concei{\c c}{\~a}o, H.~G. Silva, M.~{Collares-Pereira}, {{CSP}} mirror
  soiling characterization and modeling, Solar Energy Materials and Solar Cells
  185 (2018) 233--239.

\bibitem{heimsath_automated_2018}
A.~Heimsath, T.~Schmidt, J.~Steinmetz, C.~Reetz, M.~Schwandt, R.~Meyer,
  P.~Nitz, Automated monitoring of soiling with {{AVUS}} instrument for
  improved solar site assessment, in: {{SolarPACES}} 2017: {{International
  Conference}} on {{Concentrating Solar Power}} and {{Chemical Energy
  Systems}}, {Santiago, Chile}, 2018, p. 190008.

\bibitem{micheli_investigation_2017}
L.~Micheli, M.~Muller, An investigation of the key parameters for predicting
  {{PV}} soiling losses, Progress in Photovoltaics: Research and Applications
  25~(4) (2017) 291--307.

\bibitem{javed_modeling_2017}
W.~Javed, B.~Guo, B.~Figgis, Modeling of photovoltaic soiling loss as a
  function of environmental variables, Solar Energy 157 (2017) 397--407.

\bibitem{ballestrin_soiling_2022}
J.~Ballestr{\'i}n, J.~Polo, N.~{Mart{\'i}n-Chivelet}, J.~Barbero, E.~Carra,
  J.~{Alonso-Montesinos}, A.~Marzo, Soiling forecasting of solar plants: A
  combined heuristic approach and autoregressive model, Energy 239 (2022)
  122442.

\bibitem{bouaddi_soiled_2015}
S.~Bouaddi, A.~Ihlal, A.~{Fern{\'a}ndez-Garc{\'i}a}, Soiled {{CSP}} solar
  reflectors modeling using dynamic linear models, Solar Energy 122 (2015)
  847--863.

\bibitem{bouaddi_comparative_2017}
S.~Bouaddi, A.~Ihlal, A.~{Fern{\'a}ndez-Garc{\'i}a}, Comparative analysis of
  soiling of {{CSP}} mirror materials in arid zones, Renewable Energy 101
  (2017) 437--449.

\bibitem{bouaddi_modeling_2018}
S.~Bouaddi, A.~{Fern{\'a}ndez-Garc{\'i}a}, A.~Ihlal, R.~Ait El~Cadi,
  L.~{\'A}lvarez-Rodrigo, Modeling and simulation of the soiling dynamics of
  frequently cleaned reflectors in {{CSP}} plants, Solar Energy 166 (2018)
  422--431.

\bibitem{picotti_development_2018}
G.~Picotti, P.~Borghesani, G.~Manzolini, M.~Cholette, R.~Wang, Development and
  experimental validation of a physical model for the soiling of mirrors for
  {{CSP}} industry applications, Solar Energy 173 (2018) 1287--1305.

\bibitem{wolfertstetter_modelling_2019}
F.~Wolfertstetter, S.~Wilbert, F.~Terhag, N.~Hanrieder,
  A.~{Fernandez-Garc{\'i}a}, C.~Sansom, P.~King, L.~Zarzalejo, A.~Ghennioui,
  Modelling the soiling rate: {{Dependencies}} on meteorological parameters,
  in: {{SOLARPACES}} 2018: {{International Conference}} on {{Concentrating
  Solar Power}} and {{Chemical Energy Systems}}, {Casablanca, Morocco}, 2019,
  p. 190018.

\bibitem{lozano-santamaria_stochastic_2020}
F.~{Lozano-Santamaria}, J.~A. Luce{\~n}o, M.~Mart{\'i}n, S.~Macchietto,
  Stochastic modelling of sandstorms affecting the optimal operation and
  cleaning scheduling of air coolers in concentrated solar power plants, Energy
  213 (2020) 118861.

\bibitem{heimsath_effect_2019}
A.~Heimsath, P.~Nitz, The effect of soiling on the reflectance of solar
  reflector materials - {{Model}} for prediction of incidence angle dependent
  reflectance and attenuation due to dust deposition, Solar Energy Materials
  and Solar Cells 195 (2019) 258--268.

\bibitem{bellmann_comparative_2020}
P.~Bellmann, F.~Wolfertstetter, R.~Concei{\c c}{\~a}o, H.~G. Silva, Comparative
  modeling of optical soiling losses for {{CSP}} and {{PV}} energy systems,
  Solar Energy 197 (2020) 229--237.

\bibitem{you_temporal_2018}
S.~You, Y.~J. Lim, Y.~Dai, C.-H. Wang, On the temporal modelling of solar
  photovoltaic soiling: {{Energy}} and economic impacts in seven cities,
  Applied Energy 228 (2018) 1136--1146.

\bibitem{coello_simple_2019}
M.~Coello, L.~Boyle, Simple {{Model}} for {{Predicting Time Series Soiling}} of
  {{Photovoltaic Panels}}, IEEE Journal of Photovoltaics 9~(5) (2019)
  1382--1387.

\bibitem{fernandez-solas_estimation_2022}
A.~{Fern{\'a}ndez-Solas}, J.~{Montes-Romero}, L.~Micheli, F.~Almonacid, E.~F.
  Fern{\'a}ndez, Estimation of soiling losses in photovoltaic modules of
  different technologies through analytical methods, Energy 244 (2022) 123173.

\bibitem{wolfertstetter_novel_2012}
F.~Wolfertstetter, K.~Pottler, A.~A. Merrouni, A.~Mezrhab, R.~{Pitz-Paal}, A
  {{Novel Method}} for {{Automatic Real-Time Monitoring}} of {{Mirror Soiling
  Rates}} (2012) 10.

\bibitem{roth_effect_1980}
E.~P. Roth, R.~B. Pettit, Effect of soiling on solar mirrors and techniques
  used to maintain high reflectivity, Tech. Rep. SAND-79-2422, {Sandia National
  Lab. (SNL-NM), Albuquerque, NM (United States)} (Jun. 1980).

\bibitem{holsen_dry_1992}
T.~M. Holsen, K.~E. Noll, Dry deposition of atmospheric particles: Application
  of current models to ambient data, Environmental Science \& Technology 26~(9)
  (1992) 1807--1815.

\bibitem{ilse_fundamentals_2018}
K.~K. Ilse, B.~W. Figgis, V.~Naumann, C.~Hagendorf, J.~Bagdahn, Fundamentals of
  soiling processes on photovoltaic modules, Renewable and Sustainable Energy
  Reviews 98 (2018) 239--254.

\bibitem{seinfeld_atmospheric_1998}
J.~Seinfeld, S.~Pandis, Atmospheric Chemistry and Physics: From Air Pollution
  to Climate Change, {Wiley-Interscience}, {New York}, 1998.

\bibitem{Seinfeld2016}
J.~H. Seinfeld, S.~N. Pandis, Atmospheric {{Chemistry}} and {{Physics}} :
  {{From Air Pollution}} to {{Climate Change}}, {John Wiley \& Sons,
  Incorporated}, {New York, UNITED STATES}, 2016.

\bibitem{picotti_evaluation_2021}
G.~Picotti, R.~Simonetti, T.~Schmidt, M.~Cholette, A.~Heimsath, S.~Ernst,
  G.~Manzolini, Evaluation of reflectance measurement techniques for
  artificially soiled solar reflectors: {{Experimental}} campaign and model
  assessment, Solar Energy Materials and Solar Cells 231 (2021) 111321.

\bibitem{harris_array_2020}
C.~R. Harris, K.~J. Millman, S.~J. {van der Walt}, R.~Gommers, P.~Virtanen,
  D.~Cournapeau, E.~Wieser, J.~Taylor, S.~Berg, N.~J. Smith, R.~Kern, M.~Picus,
  S.~Hoyer, M.~H. {van Kerkwijk}, M.~Brett, A.~Haldane, J.~F. {del R{\'i}o},
  M.~Wiebe, P.~Peterson, P.~{G{\'e}rard-Marchant}, K.~Sheppard, T.~Reddy,
  W.~Weckesser, H.~Abbasi, C.~Gohlke, T.~E. Oliphant, Array programming with
  {{NumPy}}, Nature 585~(7825) (2020) 357--362.

\bibitem{virtanen_scipy_2020}
P.~Virtanen, R.~Gommers, T.~E. Oliphant, M.~Haberland, T.~Reddy, D.~Cournapeau,
  E.~Burovski, P.~Peterson, W.~Weckesser, J.~Bright, S.~J. {van der Walt},
  M.~Brett, J.~Wilson, K.~J. Millman, N.~Mayorov, A.~R.~J. Nelson, E.~Jones,
  R.~Kern, E.~Larson, C.~J. Carey, {\.I}.~Polat, Y.~Feng, E.~W. Moore,
  J.~VanderPlas, D.~Laxalde, J.~Perktold, R.~Cimrman, I.~Henriksen, E.~A.
  Quintero, C.~R. Harris, A.~M. Archibald, A.~H. Ribeiro, F.~Pedregosa, P.~{van
  Mulbregt}, {{SciPy}} 1.0: Fundamental algorithms for scientific computing in
  {{Python}}, Nature Methods 17~(3) (2020) 261--272.

\bibitem{prahl_miepython_2023}
S.~Prahl, Miepython, Zenodo (May 2023).

\end{thebibliography}
